\documentclass[manuscript, screen]{acmart}

\usepackage[nolist]{acronym}
\usepackage{natbib}
\usepackage{graphicx}
\usepackage{subcaption}
\usepackage{wrapfig}
\usepackage{hyperref}

\AtBeginDocument{%
  \providecommand\BibTeX{{%
    \normalfont B\kern-0.5em{\scshape i\kern-0.25em b}\kern-0.8em\TeX}}}

\copyrightyear{2023} 
\acmYear{2023} 
\setcopyright{acmlicensed}\acmConference[AutomotiveUI '23]{15th International Conference on Automotive User Interfaces and Interactive Vehicular Applications}{September 18--22, 2023}{Ingolstadt, Germany}
\acmBooktitle{15th International Conference on Automotive User Interfaces and Interactive Vehicular Applications (AutomotiveUI '23), September 18--22, 2023, Ingolstadt, Germany}
\acmPrice{15.00}
\acmDOI{10.1145/3580585.3607158}
\acmISBN{979-8-4007-0105-4/23/09}

\newcommand{\ints}[1]{\textcolor{teal}{ #1}}
\newcommand{\infs}[1]{\textcolor{purple}{ #1}}
\begin{document}

\begin{acronym}[]
	\acro{AI}{Artificial Intelligence}
	\acro{ACC}{Adaptive Cruise Control}
	\acro{ADAS}{Advanced Driver Assistance System}
    \acro{API}{Application Programming Interface}
	\acro{CAN}{Controller Area Network}
    \acro{CRISP-DM}{CRoss Industry Standard Process for Data Mining}
	\acro{ECU}{Electronic Control Unit}
	\acro{GDPR}{General Data Protection Regulation}
	\acro{GOMS}{Goals, Operators, Methods, Selection rules}
	\acro{HCI}{Human-Computer Interaction}
	\acro{HMI}{Human-Machine Interaction}
	\acro{HU}{Head Unit}
	\acro{IVIS}{In-Vehicle Information System}
	\acro{KLM}{Keystroke-Level Model}
	\acro{KPI}{Key Performance Indicator}
	\acro{LKA}{Lane Keeping Assist}
	\acro{LCA}{Lane Centering Assist}
    \acro{LDTM}{Lean Design Thinking Methodology}
	\acro{MGD}{Mean Glance Duration}
	\acro{OEM}{Original Equipment Manufacturer}
	\acro{OTA}{Over-The-Air}
	\acro{AOI}{Area of Interest}
	\acro{RF}{Random Forest}
	\acro{SA}{Steering Assist}
    \acro{SUS}{System Usability Scale}
	\acro{SHAP}{SHapley Additive exPlanation}
	\acro{TGD}{Total Glance Duration}
    \acro{UI}{User Interface}
	\acro{UCD}{User-Centered Design}
	\acro{UX}{User Experience}
\end{acronym}


\title[ICEBOAT]{Exploring Millions of User Interactions with ICEBOAT: Big Data Analytics for Automotive User Interfaces}






\author{Patrick Ebel}
\authornote{Both authors contributed equally to this research.}
\email{ebel@cs.uni-koeln.de}
\orcid{0000-0002-4437-2821}
\affiliation{%
	\institution{University of Cologne}
	\city{Cologne}
	\country{Germany}}

\author{Kim Julian Gülle}
\authornotemark[1]
\email{k.guelle@campus.tu-berlin.de}
\orcid{0009-0007-1284-4402}
\affiliation{%
	\institution{TU Berlin}
	\city{Berlin}
	\country{Germany}}

\author{Christoph Lingenfelder}
\email{christoph.lingenfelder@mercedes-benz.com}
\orcid{0000-0001-9417-5116}
\affiliation{%
	\institution{MBition GmbH}
	\city{Berlin}
	\country{Germany}}

\author{Andreas Vogelsang}
\email{vogelsang@cs.uni-koeln.de}
\orcid{0000-0003-1041-0815}
\affiliation{%
	\institution{University of Cologne}
	\city{Cologne}
	\country{Germany}}

\renewcommand{\shortauthors}{Ebel et al.}

\begin{abstract}
 \ac{UX} professionals need to be able to analyze large amounts of usage data on their own to make evidence-based design decisions. However, the design process for \acp{IVIS} lacks data-driven support and effective tools for visualizing and analyzing user interaction data. Therefore, we propose ICEBOAT\footnote{\textit{\textbf{ICEBOAT}}: \textbf{I}ntera\textbf{C}tive Us\textbf{E}r \textbf{B}ehavi\textbf{O}r \textbf{A}nalysis \textbf{T}ool}, an interactive visualization tool tailored to the needs of automotive \ac{UX} experts to effectively and efficiently evaluate driver interactions with \acp{IVIS}. ICEBOAT visualizes telematics data collected from production line vehicles, allowing \ac{UX} experts to perform task-specific analyses. Following a mixed methods \ac{UCD} approach, we conducted an interview study (N=4) to extract the domain specific information and interaction needs of automotive \ac{UX} experts and used a co-design approach (N=4) to develop an interactive analysis tool. Our evaluation (N=12) shows that ICEBOAT enables \ac{UX} experts to efficiently generate knowledge that facilitates data-driven design decisions. 
\end{abstract}

\maketitle

\section{Introduction}

The growing number of features of modern touchscreen-based \acp{IVIS} and the need to evaluate them with respect to the driving context~\cite{harvey.2016} makes it increasingly complex to design \acp{IVIS} that meet user needs and are safe to use. To date, the usability and distraction evaluation of \acp{IVIS} is mostly based on qualitative feedback and small-scale user studies~\cite{ebel.2020}. However, these approaches do not scale with the increasing complexity of the design task and the increasing number of features that need to be evaluated. With limited resources for user studies, practitioners often lack customer insight and must make subjective judgments instead of evidence-based design decisions. This contradicts the principles of \ac{UCD}~\cite{ISO9241-11.2018} and becomes evident when considering that the usability of infotainment systems has been the biggest source of problems for new car owners for several years~\cite{JDPower.2020, JDPower.2021,JDPower.2022}. These shortcomings lead to an increasing need for data-driven support in the automotive \ac{UX} design process~\cite{ebel.2021a}. Although modern cars are equipped with advanced telematics solutions and collect large amounts of customer usage data, \ac{UX} experts report that this data is not being used. They either do not have access to relevant data or lack the right tools to effectively visualize and efficiently analyze it~\cite{ebel.2021a}.



In order to make user-centered design decisions, automotive \ac{UX} professionals must be able to work with large amounts of data collected from customer vehicles. We argue that big data visualization tools, which automate data processing and visualization, play a critical role in this process. They allow experts to explore how customers interact with the \acp{IVIS} in real-world conditions, and thus inform decision making. These analytical tools must be developed according to the needs of domain experts, must communicate results through visualizations that serve the information needs~\cite{keim.2008}, and should keep the overhead for \ac{UX} experts low~\cite{arbesser.2017}.

Therefore, we propose ICEBOAT, an interactive visualization tool that enables automotive \ac{UX} experts to effectively and efficiently analyze driver interactions with the center stack touchscreen to evaluate \ac{UI} designs of touchscreen-based \ac{IVIS}. The tool visualizes user interaction data, driving data, and glance data, that is collected live from production line vehicles. \ac{UX} experts can specify any task they want to analyze, either by manually specifying the customer journey, or by using an interactive \ac{IVIS} emulator. ICEBOAT automatically processes the data and generates various statistics and visualizations that are based on an interview study with \ac{UX} experts and previous work by \citet{ebel.2021}. An interactive drill-down concept allows \ac{UX} experts to start wide and zoom in to analyze individual touchscreen interactions. \ac{UX} experts can compare different flows according to performance-related and distraction-related metrics such as time-on-task, number of glances, or total glance duration. 

Following a mixed methods \ac{UCD} approach and building on the work of \citet{ebel.ICEBOAT.2022, ebel.2021}, we make the following contributions:

\begin{enumerate}
    \item We present the information and interaction needs of automotive \ac{UX} experts in analyzing large amounts of customer data to evaluate touchscreen-based \acp{IVIS}.
    \item We extend the visualizations presented by \citet{ebel.2021} to the information needs of \ac{UX} experts and develop an interaction concept from task definition to user flow analysis that supports automotive \ac{UX} experts in their data analysis.
    \item We present a tool that automates the processing and visualization of touchscreen interactions, driving data, and glance data collected from customer vehicles, allowing \ac{UX} experts to interactively explore and evaluate drivers' \ac{IVIS} interactions.
    \item We evaluate the tool with industry experts (N=12) and show that ICEBOAT meets their needs and can improve the evaluation of touchscreen-based \acp{IVIS}.
\end{enumerate}

\section{Background and Related Work}

\subsection{Data in the Design and Evaluation of IVIS}
The design and evaluation of touchscreen-based \acp{IVIS} is an important factor in the overall product design process of a car. However, consumer demands often conflict with safety regulations and guidelines~\cite{NHTSA.2014, ISO15007}. This domain-specific conflict must, therefore, be considered throughout the design process of \acp{IVIS}. To create interfaces that are enjoyable and safe to use, the design and evaluation of IVISs relies heavily on questionnaires, explicit user observation, or experimental user studies~\cite{ebel.2021a}. 

However, these studies have to be designed, planned, executed, and analyzed, making them slow and expensive. As a result, they do not meet the need of automotive \ac{UX} expert to quickly and easily gain insight into customer behavior~\cite{ebel.2021a}. Using customer data to support decision making can be a competitive advantage~\cite{ahlemann.2021}. 
Live data collection from customers and continuous analysis is already standard in web and app development, and effectively used to support decision making~\cite{saura.2021, saura.2021a, guo.2022} and continuous product improvement~\cite{mattos.2017,schermann.2018}. This is different in the automotive domain, where the decision-making culture, technology, and organization have been slow to adapt~\cite{dremel.2017}. Automotive \ac{UX} professionals report that they lack the tools to access and analyze relevant data, even though modern cars collect large amounts of driving and interaction-related data. To gain insights from customer data, \ac{UX} experts often have to submit requests to data scientists and involve other departments~\cite{ebel.2020}. As a result, the problem that traditional methods are slow is only shifted, not solved. \ac{UX} experts need to be empowered to analyze and visualize user interaction data. Developing data analysis tools that meet their needs and facilitate their design activities is critical to making their jobs easier.

\subsection{Creating Meaningful Interactions with Big Data}

In today's product development, decision makers, regardless of the domain, aim to make data-driven decisions. This allows them to design products that are tailored to customer needs~\cite{fisher.2012}. However, the main challenge in data-driven decision making is not the acquisition of raw data itself, it's the challenge of extracting useful knowledge from it~\cite{cui.2019}. To create solutions that help designers, engineers, or scientists, the right information must be available at the right time~\cite{keim.2008}. Therefore, tools and analytical solutions need to communicate the results of analysis through meaningful visualizations and clear representations~\cite{keim.2008}. 

However, creating meaningful visualizations and intuitive tools for analyzing big data is far from obvious. Although several commercial general-purpose tools exist, they often fail to meet domain and task-specific needs. Domain experts often require advanced visualization and interaction concepts that are not supported by commercially available tools. These tools stick to a small set of standardized visualizations~\cite{zhang.2012}.

While these standardized dashboards are a valuable tool for quickly communicating \acp{KPI} to stakeholders or managers, they are often disconnected from the domain expert's workflow and serve as a reporting tool rather than an exploratory knowledge generation tool. To support domain experts in their work, it is important to create solutions that are specific to their workflow. Individual visualizations often address a specific task that is part of a larger workflow. Therefore, these visualizations need to be linked in such a way that they support this workflow as a whole~\cite{arbesser.2017}. An example of this is the common need to explore large amounts of data at multiple scales. This can be achieved by visualizing the data at different levels of granularity, starting broadly and zooming in on details as the analysis progresses~\cite{fisher.2012}. In addition, domain experts are often non-specialists when it comes to analyzing large amounts of data. It is therefore important to avoid information overload and to use visualizations that are easy to understand and whose benefits are immediately apparent.

\subsection{Big Data Visualizations to evaluate Automotive User Interfaces}

The evaluation of \acp{IVIS} differs from web or mobile applications. While traditional usability metrics such as time on task or error rates play an important role, they are far from sufficient for evaluating a driver's interaction behavior holistically. Drivers often interact with \acp{IVIS} while driving even though they are required to constantly monitor the driving scene even in partially automated driving as we see it on the road today (Level 1 and 2 according to SAE~\cite{SAE.2021}).

This makes it important to not only evaluate the usability~\cite{ISO9241-11.2018} of touchscreen-based \acp{IVIS} but also their distraction potential. Regarding the assessment of distraction, the visual demand of interfaces has proven to be an effective measure, as long glances away from the road (\textit{t>2\,s}) are directly correlated with increased crash risk~\cite{Klauer.2006}. As a result, the evaluation of \ac{IVIS} in terms of usability and driver distraction is a well-researched topic~\cite{harvey.2011, harvey.2016, frison.2019, ebel.2023, grahn.2020}. However, there is a considerable gap between the academic research conducted to evaluate \acp{IVIS} and the tools and methods available to the professionals in industry who eventually design these systems. Even though automotive \ac{UX} professionals express clear needs for effective visualizations to support their work, there is not much research on big data analytics to evaluate user interactions with IVIS. Most visual analytics approaches in the automotive domain focus on visualizing data collected from a few sensors~\cite{sedlmair.2011} or in controlled experimental studies~\cite{jansen.2023}. For example, \citet{jansen.2023} present an approach to visualize spatiotemporal data collected during user interface interaction studies. Their approach provides valuable insights into the combined visualization of explicit and implicit data from different sources. However, the tool is built according to academic needs and focuses on the visualization of individual situations recorded during user studies. Therefore, it is not applicable to the challenges faced by industrial \ac{UX} experts when they need to analyze millions of data points collected live from customers.  In this context, driving, glance, and interaction data collected live and in large quantities from customer vehicles \ac{OTA} must be automatically processed, stored, and visualized to allow \ac{UX} experts to evaluate the usability and distraction potential of \ac{IVIS}. Considering the automotive specific requirements and constraints that apply to the analysis and visualization of automotive event sequence data~\cite{ebel.2021, jansen.2023, ebel.2021a}, most related approaches~\cite{wongsuphasawat.2012,matejka.2013,liu.2017, crosslin.2021} and commercial alternatives (e.g., UserTesting\footnote{\url{https://www.usertesting.com}}, UserZoom\footnote{\url{https://www.userzoom.com}}) are not suited to the problem at hand.  A solution explicitly focused on industry professionals has been proposed by~\citet{ebel.2021}, who have developed an approach to visualize large amounts of user interaction, driving, and glance behavior date. They visualize this data at three levels of granularity and show that \ac{UX} experts can use these visualizations to evaluate secondary touchscreen interactions. However, they only propose prototypes of these visualizations and do not provide insight into whether the visualizations meet the needs of \ac{UX} professionals and how \ac{UX} experts should use these visualizations in their workflow.

\section{Approach}


Our approach aims to improve the industrial design process of \acp{IVIS}. Designers and \ac{UX} researchers report that they are often forced to neglect design evaluation due to time constraints and data accessibility issues~\cite{ebel.2020}. To develop a solution that meets the needs of automotive \ac{UX} experts and improves their design and evaluation process, we followed a mixed methods \acf{UCD} approach~\cite{still.2017, abras.2004}.
In the first phase, we conducted semi-structured background interviews (n=4) to extract requirements according to information and interaction needs and to evaluate the visualizations originally proposed by \citet{ebel.2021}. Using a participatory design approach, we then co-designed prototypes with four automotive \ac{UX} experts. Throughout the co-design approach, we walked different focus groups through the current state of the prototypes and discussed potential improvements and necessary changes.
After the fourth co-design session, we evaluated the prototype in a second study where we conducted usability testing and explored the context of use of the prototype.

\section{Study 1: Extracting Information and Interaction Needs}

The first study has two objectives: First, to confirm the results presented by \citet{ebel.2021}, who claim that their multi-level user behavior framework was found useful by automotive \ac{UX} experts. Second, \citet{ebel.2021} propose three visualizations, but do not present any insights or solutions on how these visualizations can be connected to effectively support \ac{UX} experts in evaluating \acp{IVIS}. Therefore, we extract detailed information and interaction needs that form the basis for the subsequent co-design process. For this purpose we conducted semi-structured interviews.

\subsection{Participants}

We conducted semi-structured interviews with 4 \ac{UX} experts (I1 - I4). All of them had between five and nine years of professional experience. At least five of those years were spent in the field of \ac{UI} design. They have also all been with Mercedes-Benz for more than five years. We therefore consider them to be knowledgeable about automotive design processes and working methods.

\subsection{Procedure}
The interview agenda consisted of three parts: \textit{introduction}, \textit{main section}, and a \textit{conclusion}. Following the recommendations of~\citet{renner.2020}, we prepared a set of open-ended questions and optional follow-up questions for each section. The latter were intended to refine ambiguous responses and further guide the interview. In addition, we periodically summarized the responses during the interview to reflect and confirm correct understanding. While the introduction was designed to create an open atmosphere and establish a common ground, we ended each interview with a conclusion, asking the interviewees if there was anything they wished to add. The main section contained the majority of the questions. Here we asked the interviewees about their \textit{information needs}, \textit{interaction needs}, and the \textit{visualizations} they would expect to see in a potential tool that supports their current workflow. Regarding the visualizations, we gave the interviewees some time to develop their ideas. We then presented the interviewees with the aforementioned visualizations suggested by~\cite{ebel.2021}. By showing them the existing visualizations, we hoped to support their ideation process~\cite{still.2017} and wanted to confirm the informal evaluation presented by \citet{ebel.2021}.

\subsection{Information Needs}

After analyzing the coded interview transcripts, we identified 39 information needs that fall into 7 categories. We present these needs below, where \textit{Count} refers to the number of unique needs within a category and \textit{Support} refers to the number of total needs expressed by the participants.

\textbf{\infs{INF-1}: Usability and Distraction-Related Metrics} (\textit{Count = 8, Support = 12}).
Driver interactions with \acp{IVIS} while driving are considered secondary tasks~\cite{regan.2022}. Thus, not only usability but also driver distraction play a major role in evaluating automotive user interface~\cite{ebel.2021a}. Accordingly, the respondents formulated various information needs that revolve around the understandability of the \ac{UI} (I1, I2, I3), performance-related metrics such as time on task (I1), error rates (I1, I2), or the number of interactions needed to perform a task (I2). They also stated that for a holistic evaluation they need to be able to evaluate the visual demand (e.g. number of glances) of features (I1, I2) and individual user flows (I1). They also stressed the importance of being able to see the correlation between \ac{IVIS} usage and driving data.

\textbf{\infs{INF-2}: Feature Usage Information} (\textit{Count = 8, Support = 12}).
The automotive industry is moving from a technology-driven development approach to a more user-centric one~\cite{bryant.2014}. While this process has been going on for many years, there is still a lack of knowledge about how features are used by customers. This leads to many features being carried over from old releases that may not be needed by customers~\cite{ebel.2021a}. During the interviews, feature usage information was the first \ac{KPI} that respondents thought of. The typical questions UX professionals want to answer based on data insights are questions like \textit{``How often is a feature used?''} (I1-I4) or \textit{``How long is a feature used on average?''} (I1, I2). Participants indicate that information about feature usage is valuable because it is often used to decide whether to continue or discontinue a feature.

\textbf{\infs{INF-3}: Usage Pattern Visualizations} (\textit{Count = 7, Support = 11}).
To gain deeper insights into user behavior, \ac{UX} experts expressed different needs regarding the analysis of user flows and how users interact within certain features (\textit{Usage Patterns}). They want to know how people use the system (I2, I4), how they navigate the system to perform certain tasks (I1, I4), and what kind of \ac{UI} elements they use (I2). Participants are also interested in merging this information with usability and distraction-related metrics (e.g., to compare different flows).

\textbf{\infs{INF-4}: System Information} (\textit{Count = 6, Support = 10}).
The cars in an \ac{OEM}'s fleet are very heterogeneous, both in terms of hardware and software. Not only do manufacturers offer different models that differ according to the market in which they are sold, but customers can also configure their cars according to their personal preferences (e.g., different sizes of center stack touchscreens)~\cite{broy.2007}. This, combined with the long product lifecycle and limited ability to perform \ac{OTA} updates, especially for older models, results in many different \ac{UI} versions being used by customers. This is reflected in the information needs of \ac{UX} professionals. They state that they need to compare usability and distraction-related metrics, feature usage information, and usage patterns across car models (I1-I4), software versions (I2, I4), screens (driver vs. front passenger vs. rear passengers), and screen sizes (I2). This information is needed to assess the interplay between hardware and software but also to track progress.

\textbf{\infs{INF-5}: Contextual Information} (\textit{Count = 5, Support = 6}).
Driver behavior and driver interactions are highly context sensitive~\cite{ebel.2023} and participants state that they need contextual information to better judge individual interaction sequences. For example, they state that they need information about the driving situation (1, 3) to be able to judge how drivers interact in different situations. They also want to know how many passengers were present (2) and whether a cell phone was connected to the \ac{IVIS} (2), arguing that these could be additional sources that influence driver behavior without being represented in the interaction, glance, or driving data.

\textbf{\infs{INF-6}: Input Modalities} (\textit{Count = 3, Support = 4}).
Participants were also interested in the different types of modalities that drivers or passengers can choose to interact with \ac{IVIS} (e.g., different modes of touch interaction, voice, or steering wheel control). In particular, they want to know which modality drivers primarily use (1) and whether this use differs across features (2,4).

\textbf{\infs{INF-7}: User Information} (\textit{Count = 2, Support = 3}).
For user-specific information, respondents see value in comparing data from different regions (3, 4) or comparing data for different target groups (e.g., by demographics or frequently used features).

Regarding the visualizations proposed by \citet{ebel.2021}, participants agreed that they already partially address the information needs \infs{INF-1}, \infs{INF-2}, \infs{INF-3}, and \infs{INF-5}. However, they do not provide system information (\infs{INF-4}), information about different modalities (\infs{INF-6}), or user information (\infs{INF-7}).

\subsection{Interaction Needs}

To extract the interaction needs of the participants, we asked them to imagine a tool that would meet all their information needs and to explain how they would like to use this tool in their daily work. The expectations were very consistent, as they all expected to use the tool to define a new \ac{UI} concept, to validate an existing and already implemented \ac{UI} concept, and to question the customer value of a feature. Based on these insights, we then explored how users would like to interact with the anticipated tool and how they would like to configure it to meet their needs. The answers to these questions form the interaction needs. As shown below, we grouped the 14 individual needs into 4 categories.

\textbf{\ints{INT-1}: Task Definition} (\textit{Count = 4, Support = 10}).
Participants emphasized that they want to configure their analytics based on individual use cases, rather than having a "one-size-fits-all" dashboard. While they valued certain standard metrics to be displayed, they wanted to define specific tasks or characteristics for which they needed detailed analytics. To define the tasks of interest, all participants (I1-I4) asked if it would be possible to interactively define sequences without having to manually enter the object identifiers. They suggested using a desktop-based version of \ac{IVIS}, arguing that this would facilitate task definition since the \ac{UI} software consists of thousands of elements. However, for known use cases, they suggested traditional input options such as drop-down menus to select \ac{UI} elements as start and end points (I1, I2). Here, one participant (I2) mentioned that the analysis tool should use the same \ac{UI} identifiers as those used in the \ac{UI} concept description.

\textbf{\ints{INT-2}: Analysis} (\textit{Count = 5, Support = 13}).
When it came to analyzing, participants were concerned about overall complexity, noting that traditional dashboards often tend to be overloaded and cluttered. Accordingly, they asked for features that would allow them to reduce the complexity of the results. They also wanted to be able to drill down through different levels of granularity depending on their use case, rather than being presented with all the results at once (I1, I2, I4). All participants argued that they need to be able to compare usage by system, context, and user information (I1-I4). Most of the proposed filtering options focused on system-specific information such as car type or software version.

\textbf{\ints{INT-3}: Operating Aids} (\textit{Count = 3, Support = 5}).
Two participants (I1, I2) mentioned that the tool should be adaptable according to the user's expertise. They suggested that the tool could provide an ``exploration mode'' (I2) to help them explore the \ac{UI}. They also asked for the possibility to display reduced versions of the plots proposed by~\cite{ebel.2021}.

\textbf{\ints{INT-4}: Sharing and Collaboration} (\textit{Count = 2, Support = 4}).
Participants expressed the need to share the visualization with colleagues and decision makers, either in a portable format (I1-I3) or through a link that provides direct access (I3).

The visualizations presented by \citet{ebel.2021} are stand-alone visualizations without a user interface. Therefore, they do not address any of the identified interaction needs.

\section{Introducing ICEBOAT}

	\begin{figure*}[tbp!]
		\centering
		\includegraphics[width=\linewidth]{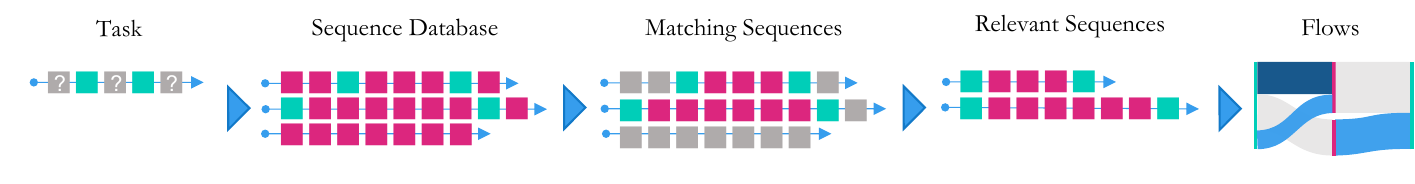}
		\caption{Sequence and flow extraction according to task definition. Based on the first and last interaction of the defined task (green squares), all sequences in the database that match the requirements are extracted. All remaining sequences (relevant sequences) are then aggregated to generate the flow statistics.}
		\label{fig:Task_to_Flow}
	\end{figure*}

Study 1 identified the information and interaction needs of automotive \ac{UX} experts for visualization and analysis of customer data and confirmed that the visualizations presented by \citet{ebel.2021} partially satisfy the information needs of \ac{UX} experts. However, they do not provide an interface that addresses the interaction needs. Therefore, following \infs{INF-1 -- INF-7} and \ints{INT-1 -- INT-4}, we developed ICEBOAT, an interactive user behavior analysis tool for automotive \acp{UI}. ICEBOAT refines the visualizations of \citet{ebel.2021}, adds new functionalities and connects them in a meaningful way. Built on top of the telematics data logging framework introduced by~\cite{ebel.2021}, it automates task definition, data processing, and visualization generation, making large amounts of customer data easily accessible for \ac{UI} evaluation.

We developed ICEBOAT using a co-design approach with four iterations. We invited the background interview participants as co-designers to each of the sessions, which were conducted remotely using Microsoft Teams. 

\subsection{System Architecture}

ICEBOAT consists of a web-based frontend application for data visualization and a backend system for data processing (see Appendix \ref{app:SequenceDiagram}). The frontend, developed using the JavaScript framework \textit{Vue.js}\footnote{\url{https://vuejs.org}}, receives data from three different services: The \textit{Concept Database} (containing all \ac{UI} information), the \textit{IVIS Emulator} and the \textit{Backend}. The \textit{IVIS Emulator} virtualizes the \ac{IVIS} so that it can be executed on a computer as if it were running in the car.

The backend is divided into two services: An \ac{API} service built with \textit{FastApi}\footnote{\url{https://fastapi.tiangolo.com/}} web framework and a data service. The API service receives the analysis requests, passes them to the data service, and returns the results. The data service uses \textit{PySpark}\footnote{\url{https://spark.apache.org/docs/latest/api/python/index.html}} to efficiently extract, transform, and load the customer data stored in the data lake. The data lake is updated daily with the latest customer data. After running the analytical queries and extracting relevant user flows (see \autoref{fig:Task_to_Flow}), the backend returns the results to the API service. The frontend enhances the processed data with additional UI-specific information from the \textit{Concept Database}. We chose this architecture to make ICEBOAT easily extensible and to ensure interoperability (e.g. with another back end solution)~\cite{sadeghi.2023}.

	\begin{figure*}[tbp!]
		\centering
		\includegraphics[width=\linewidth]{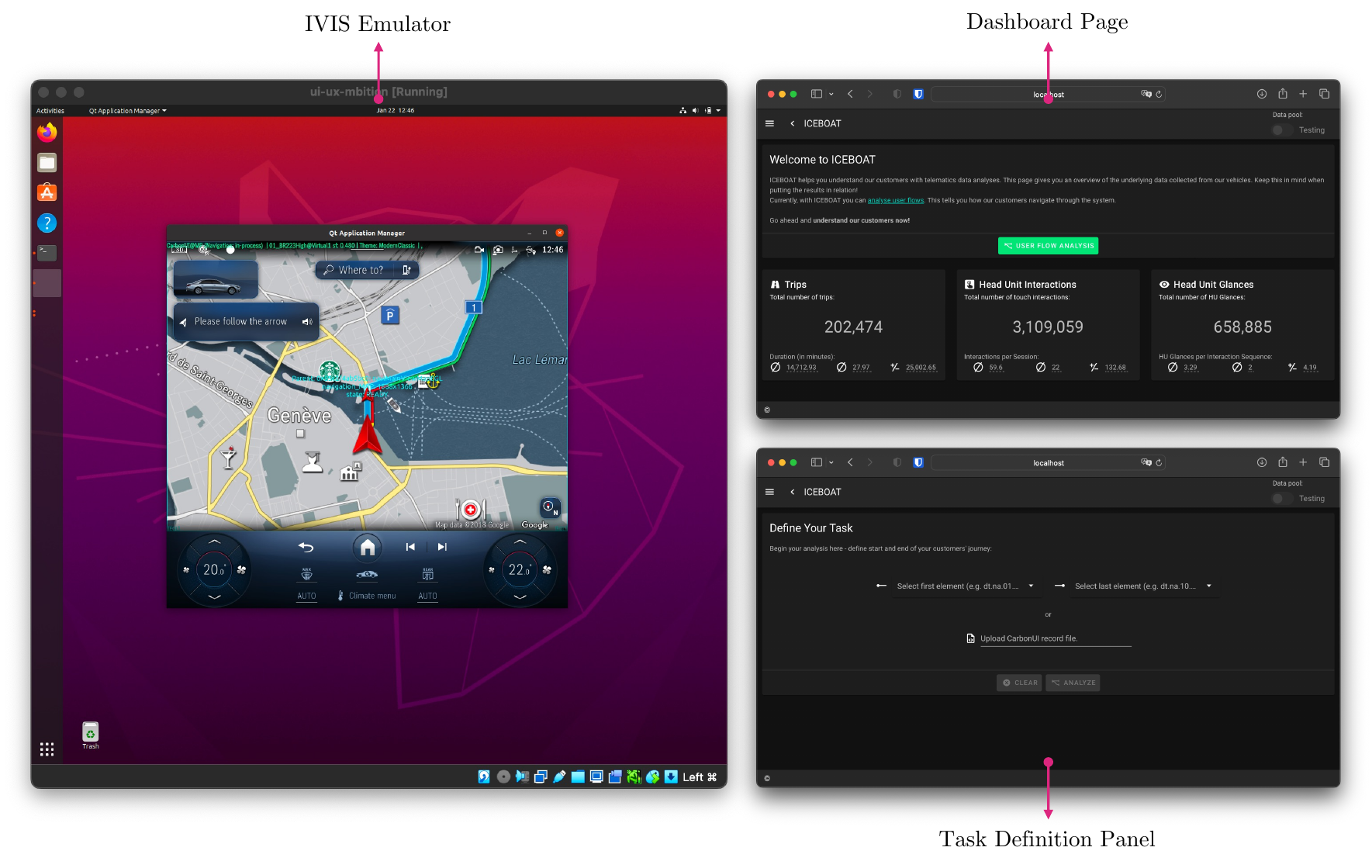}
		\caption{Overview of the \ac{IVIS} Emulator (left), the Dashboard View (top right) and the Task Definition Window (bottom right)}
		\label{fig:Emulator_Dashboard_Input}
	\end{figure*}

\subsection{Interactive Web Application and IVIS Emulator}

 	\begin{figure*}[tbp!]
		\centering
		\includegraphics[width=\linewidth]{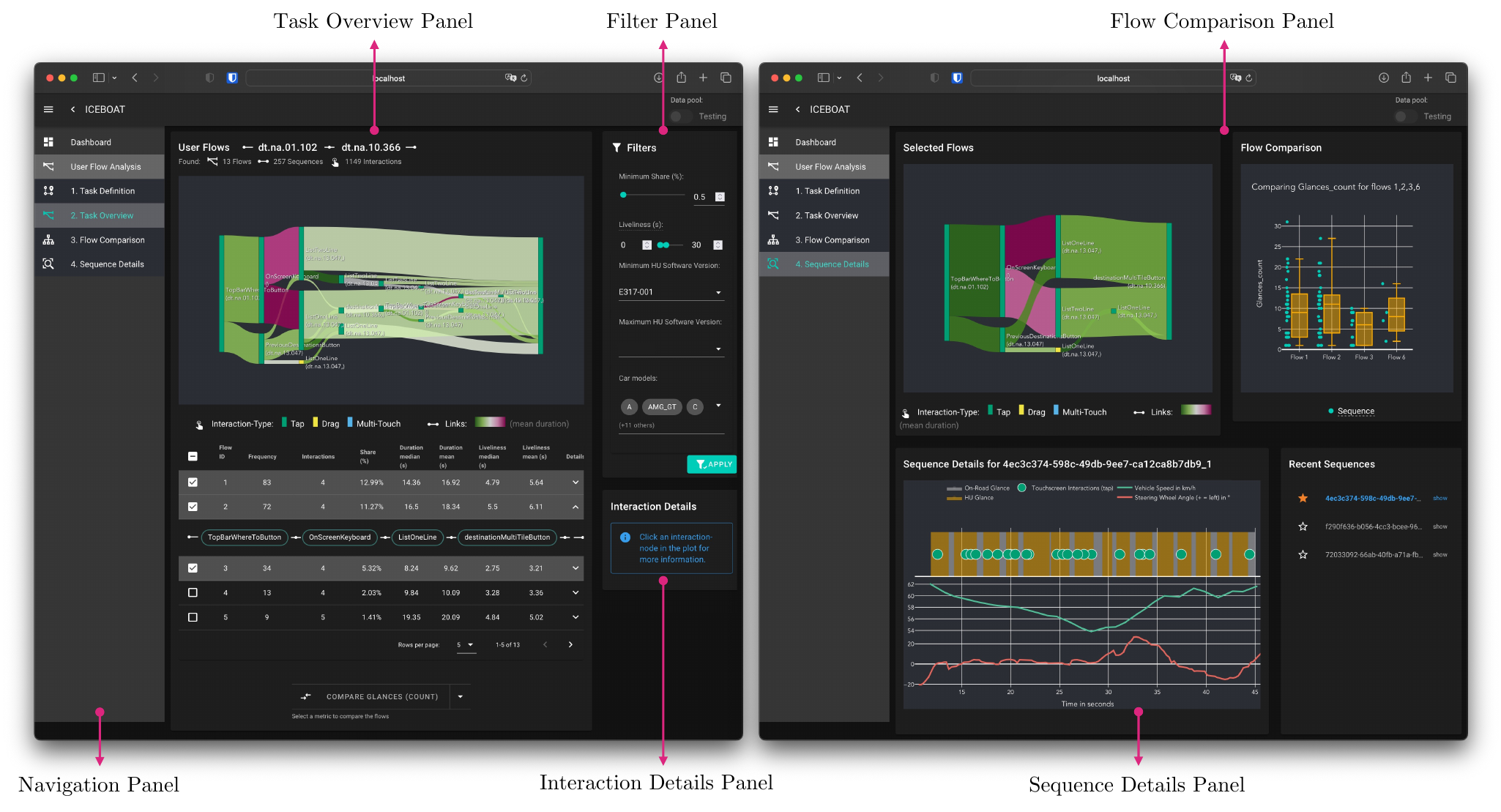}
		\caption{Overview of the User Flow Analysis Page The left part shows the initial Task Overview Panel. The Sankey diagram consists of all flows that satisfy the task and filter requirements. The table below the Sankey diagram shows various flow-related user behavior metrics. Clicking a UI element in the Sankey diagram opens the Interaction Details Panel, which displays the technical description of the element. On the right is the Flow Comparison panel, which consists of a reduced Sankey diagram and box plots showing the distribution of the total number of glances per interaction sequence. The Sequence Details Panel shows the touch interactions, gaze, and driving behavior of a specific sequence over time. The given example represents the use case of Study 2.}
		\label{fig:Analysis_View}
	\end{figure*}

Figures~\ref{fig:Emulator_Dashboard_Input} and \ref{fig:Analysis_View} show an overview of the final tool after four iterations. The interface  consists of a \textit{Dashboard Page} and a \textit{User Flow Analysis Page}. The user flow analysis page consists of four panels that fade in one after another based on the user's input addressing the need for a drill down mechanism to reduce complexity (\ints{INT-1}). An overview of all components is given below:

\begin{enumerate}
    \item \textbf{Dashboard Page:} Upon opening ICEBOAT, users are presented with a dashboard page (see Figure~\ref{fig:Emulator_Dashboard_Input}) that welcomes them and gets them started by explaining the purpose of the application. The tiles below the introduction report specific \acp{KPI} that describe the underlying data. For example, the number of trips on which the analysis is based and the number of logged interactions with the head unit. This page therefore onboards users and provides a perspective on the data to facilitate entry into the analysis (\ints{INT-3}).
    \item \textbf{Task Definition:} The tool provides two ways for users to define the task they want to analyze. (1) They can select the \ac{UI} elements that define the start and end of a task from a searchable drop-down menu filled with all the \ac{UI} elements that exist within the \ac{IVIS}. 
    (2) Alternatively, you can define a task using the \textit{IVIS Emulator} (see Figure~\ref{fig:Emulator_Dashboard_Input}). ICEBOAT then automatically extracts all similar flows from the customer data and visualizes them. This provides the user with a playful and easy way to define tasks without having to know the naming conventions of specific \ac{UI} elements. These two options address the interaction needs of \ints{INT-1} and \ints{INT-3}.
    
    \item \textbf{Task Overview:} After loading the data for the specified task, the \textit{Task Overview Panel} (see Figure~\ref{fig:Analysis_View}) presents aggregations for all user flows between the start and end event of the task (\infs{INF-3}). The data is presented as an adapted Sankey diagram~\cite{ebel.2021} and in tabular form. The two views provide information about the average time between two consecutive interactions in a flow, the gestures used, the relative and absolute frequency of flows, the total number of interactions in a flow, and the average flow duration (\infs{INF-1, INF-3}). The \textit{Filter Panel} on the right-hand side allows users to customize the visualization to reduce visual clutter (\ints{INT-2}). It also allows filtering for specific software versions or car types (\infs{INF-4}). The visualized flows are based on all of the user interaction sequences collecte from production vehicles that match the task definition (see Figure~\ref{fig:Task_to_Flow}) and the applied filters.
    
    \item \textbf{Flow Comparison:} The \textit{Flow Comparison Panel} shows a reduced Sankey diagram of the selected flows (\ints{INT-3}) and a box plot comparing each of the flows according to a selected metric, such as gaze duration or time on task (\infs{INF-1}, \infs{INF-3}). The box plots show the distribution of these values for all sequences contained in the flow. Thus, each dot represents a single sequence, which users can select (by clicking) to open the Sequence Details view for that interaction sequence.
    \item \textbf{Sequence Details:} The \textit{Sequence Details Panel} allows the user to explore details about a single interaction sequence. The view blends glance data (on-road, off-road, center stack touchscreen) with contextual driving data such as speed or steering angle, and embeds the touchscreen interactions~\cite{ebel.2021} (\infs{INF-5}). On the right, users see a history of sequences they have viewed and can save specific ones as favorites (\ints{INT-3}).
\end{enumerate}

\subsection{How ICEBOAT Empowers Automotive UX Experts}

The provided visualizations and analyses support UX experts in the design and evaluation of touchscreen-based \ac{IVIS}. With ICEBOAT, we support professionals in overcoming three key challenges related to the use of big data analytics in the design and evaluation of IVIS:

\textbf{1. Data-Driven Decision Making:} 
Due to cultural, organizational, and technological challenges, data plays only a minor role in the decision-making process related to automotive design~\cite{dremel.2017}. As technology improves, large amounts of driving- and interaction-related data are being collected. However, \ac{UX} professionals still report that they lack the tools to access and analyze the data directly and independently~\cite{ebel.2021a}. 
Based on a telematics data processing framework, ICEBOAT gives \ac{UX} professionals permanent and immediate access to usage data collected live in the field. This allows \ac{UX} experts to analyze interaction, glance and driving data independently throughout their workflow.

\textbf{2. Automotive-Specific Analysis:} General-purpose tools for big data analytics are often disconnected from the workflows of domain experts~\cite{arbesser.2017}. This is also true in the automotive domain. \ac{UX} experts report (Study 1) that current tools do not meet their specific needs for task definition, user flow exploration and comparison, and visualization of individual usage sequences. ICEBOAT supports the definition of user tasks, allows users to compare specific flows according to various usability and driver distraction related metrics, and enables \ac{UX} experts to visualize details of individual interactions with the \acp{IVIS}.

\textbf{3. Information Overload:} \ac{UX} experts are not trained to analyze large amounts of data. Therefore, data visualizations and interactions must be easy to understand and their benefits obvious to avoid information overload~\cite{keim.2008}. ICEBOAT allows \ac{UX} experts to explore UI-relevant data without requiring technical knowledge. The \ac{IVIS} emulator allows \ac{UX} experts to easily define the scope of their analysis and pre-defined \acp{KPI} are visualized decoupled from the detailed user flow analysis to avoid information overload. The user flow analysis allows users to start with a broad overview and then zoom in on details as the analysis progresses. This drill-down concept (panels appear one after the other) fits the workflow of \ac{UX} experts and presents only the information that is needed.

\section{Study 2: Evaluation with Designers and Data Scientists}
To assess whether ICEBOAT enables \ac{UX} experts to independently explore large-scale behavioral data, we conducted an evaluation study with \ac{UX} researchers, designers, and data scientists.

\subsection{Method}
We conducted usability testing, interviews, and a context of use questionnaire. We aimed for a representative sample of participants and used standardized measures to assess usability. To guide the evaluation, we followed a test plan that we created based on the recommendations of~\citet{still.2017}.

\subsubsection{Participants}
We recruited 12 potential users from Mercedes-Benz and MBition, a Mercedes-Benz software hub: 4 designers, 4 \ac{UX} researchers, and 4 data scientists. We included data scientists for two reasons: First, due to cross-functional development teams, data scientists often work closely with designers or \ac{UX} researchers in decision-making processes. Second, because of their familiarity with data analysis, we expected data scientists to provide a different perspective and baseline for understanding data. We did not invite participants who had already participated in the study, as this could skew the results when evaluating usability~\cite{mclellan.2012,albers.2020,borsci.2015}. The age of the participants ranged from 21 to 41 years (mean 29.6, SD 5.6) and their work experience from 0.5 to 20 years (mean 5.3, SD 5.8). All but one participant had a college degree. 


\subsubsection{Scenario and Evaluation Tasks}
To create a realistic evaluation environment, we derived a test scenario from the storyboard resulting from Study 1 (see Figure \ref{fig:StoryBoard} in \ref{app:Storyboard}). In this scenario, the \ac{UX} experts are asked to evaluate the destination entry task of the navigation feature. They should use the \ac{IVIS} Emulator to define a representative task and then analyze it for bottlenecks, driver distraction, and outliers in glance behavior. 

\subsubsection{Procedure}
We collected demographic data in a pre-survey before the experiment. At the beginning of the experiment, we introduced the scenario and asked the participants to complete seven evaluation tasks (compare~\ref{app:Tasks}) that resembled the scenario introduced above. First, we shared the \ac{IVIS} emulator with the participants and asked them to complete the first task. Then we switched to the ICEBOAT screen. Before the next task, we had the participants practice thinking aloud by asking them to describe the dashboard page and give feedback. Users then navigated to the \textit{User Flow Analysis} page of the tool and proceeded with the second task. During the test, participants were free to explore the tool and ask questions. After the participants completed all tasks, we collected their feedback both verbally and with the post-survey. We also recorded whether participants encountered any technical problems during the test. Since participants in a lab study usually answer usability questionnaires on-site~\cite{still.2017}, we had the participants fill out the surveys online immediately after the study. However, we stopped recording the interviews and turned off the cameras and microphones so that participants would not feel observed while completing the survey.

\subsubsection{Measures}
We counted and coded the errors participants made while solving the tasks, and collected participants' feedback and interpretations of the visualization. We also had the participants fill out the \ac{SUS} questionnaire~\cite{brooke.1996} and the Context of Use questionnaire (see \ref{app:ContextofUse}).

\subsubsection{Test Environment \& Schedule}
Due to the distributed work environment, we conducted all experiments remotely using Zoom. With the users' permission, we recorded each test session to analyze the session afterwards and to quantify the error rates per task. We prepared a setup with the \ac{IVIS} emulator open on one screen and the ICEBOAT tool open on another. This mimics the setup we imagine users would have when using the prototype in production. Using screen sharing, we allowed participants to remotely interact with the emulator and analysis tool. This makes our test environment as similar as possible to the production environment. We ran 12 tests of one hour each over three weeks.

\subsection{Quantitative Results}
We present results from the \ac{SUS} and Context of Use questionnaires, as well as additional qualitative insights.

\subsubsection{SUS}
ICEBOAT received a mean \ac{SUS} score of 68.125 (MD=70, SD=16.89), which, according to \citet{lewis.2018}, is average. While data scientists rated the tool with a mean score of 80, \ac{UX} experts rated it with a mean score of 62. \citet{cleland.2019} reports a similar spread between domain experts and data scientists. When evaluating the usability of the proposed big data analytics platform, data science testers rated the platform almost 20 points higher than policymakers (75.0 vs. 56.7).

\subsubsection{Context of Use}
The mean score of the Context of Use questionnaire was 4.2 out of 5 (MD=4.24, SD=0.33). In contrast to the SUS, only two questions were rated differently by data scientists and \ac{UX} experts. The Pearson correlation between the results of the Context of Use and SUS questionnaires was not significant (R=0.55, p=0.061), suggesting that usability and value to the experts' workflow are not directly related.

\subsection{Qualitative Feedback}

Overall, participants found the tool valuable and easy to use. They reported that it would open up new possibilities for them and make their workflow much more efficient, \textit{``I think this really makes our job easier, especially when you see how quickly you can get evaluations compared to how long it takes now.''} (P3) (\ints{INT-2}). They also report that ICEBOAT provides effective insights because it \textit{``[...] would provide better answers to many questions''} (P9). P9 further states: ``It is relatively difficult for us to make statements about groups that drive premium vehicles. With the tool you could get to those people.'' (\infs{INF-7}). They also appreciated the ability to define the task using the \ac{IVIS} Emulator, as it allows them to define the scope of their analysis without having to know the identifiers of specific \ac{UI} elements (\ints{INT-1}). They report that this facilitates exploration and reduces the burden of using this tool. The design and layout of the tool was generally well received.

\subsection{Data Understanding}

ICEBOAT effectively stimulated discussion about usability and safety improvements as participants solved the tasks. The Sankey diagram visualization was easy for participants to understand, and they were able to identify bottlenecks using the color scale or by manually comparing interaction times (shown when hovering over the flows) (\infs{INF-3}). One participant immediately suggested that the search suggestions could be improved to reduce the number of characters the driver has to type, because \textit{``[t]he list keeps updating as you type. So it takes the user more time to find what they are looking for if they type more characters.''} (P8).
When comparing the top three flows based on the number of glances, 5 participants asked for clarification on how to interpret the box plots, but were able to identify the flow with the lowest average number of glances once explained (\infs{INF-1}). Participants quickly identified the flow with the most glances (\infs{INF-1}) and appreciated the ability to select individual sequences to open the Sequence Details Panel (\ints{INT-2}). Using the Sequence Details Panel, they were able to assess the dependencies between glance, interaction, and driving behavior (\infs{INF-5}), \textit{``The driver is on the move and slows down in the course of the interaction''} (P6), \textit{``after brief glances at the road, the driver immediately performs several interactions''} (P10). However, participants interpreted the steering angle changes differently, with some interpreting them as a sign of distraction and others as a driving maneuver. Overall, participants found the tool helpful and argued that the insights can be particularly valuable in defining the scope of specific user studies to explore not only the ``what'' but also the ``why''.

\subsubsection{Errors}
In general, participants reported that they understood the tasks easily and were able to complete them efficiently. 
When interacting with ICEBOAT (Task 2-7), participants made only minor errors. For example, two participants initially chose a minimum support that was too high or too low, making the visualization either too cluttered or too sparse. Also, to create a reduced Sankey diagram in Task 5, four participants wanted to further reduce the flows using the minimum support instead of using the checkboxes in the table.
Most of the errors occurred when interacting with the \ac{IVIS} Emulator. While all participants successfully created a recording, only five out of twelve users did so with the expected start and end, as they did not start the recording on the expected screen. When asked, the participants stated that they thought they should start the recording directly from the main menu. However, this is more of a study-induced error with no practical implications.
When asked to elaborate on their errors, participants stated that it takes some time to get used to the tool but \textit{``[o]nce you get used to it and it's established as a working tool, it's super helpful.''} (P12).

\section{Limitations and Future Work}

While our results show that ICEBOAT effectively empowers \ac{UX} experts and meets most of the information and interaction needs for analyzing large amounts of interaction data, some limitations should be considered. First, data scientists rated the usability of the tool higher than \ac{UX} experts. This may be due to their experience with other data analysis tools. However, it also suggests that further research should be conducted to address the shortcomings with respect to users unfamiliar with data analysis. In addition, the slight delay and minor issues with the screen sharing and remote control feature may have influenced the results.
Second, the study only considered touch interactions on the center stack screen. However, drivers can also interact with \acp{IVIS} using speech or hardkeys. Thus, to satisfy \infs{INF-6}, the next step would be to introduce these modalities in ICEBOAT.
Furthermore, we only interviewed employees of one \ac{OEM}. While related work~\cite{ebel.2021a, dremel.2017} suggests that development practices and challenges are similar across most automotive \acp{OEM}, information and interaction needs may be skewed.
Due to privacy concerns, we are not allowed to collect personal data. Thus, the only way to satisfy \infs{INF-7} is to use a combination of available filters to define ``target groups'' (e.g., luxury car buyers vs. compact car buyers, as indicated by P9).
Finally, we recorded the tests remotely, and 4 participants reported that the remote control function temporarily stopped working. While we were able to immediately restore control for 3 of the 4 people, this prevented one participant from completing a task. We had this participant verbally instruct us to complete the task and then restored remote control. 

\section{Conclusion}

We present ICEBOAT, an interactive tool that makes millions of in-vehicle user interactions available to \ac{UX} experts to effectively and efficiently visualize and evaluate drivers' touchscreen interactions with \acp{IVIS}. 

In Study 1, we identify the information and interaction needs of \ac{UX} experts when analyzing large amounts of telematics data. Our findings reveal a design trade-off: \ac{UX} experts want to access as much data as possible and perform IVIS-specific analyses, but are deterred by the complexity of traditional big data visualization tools. ICEBOAT addresses this conflict of interest by (1) allowing users to define a task via a \ac{IVIS} emulator, (2) automating all data processing and cleaning while still allowing manipulation of the metrics that matter, and (3) providing an interactive drill-down mechanism that allows users to start broad and zoom into the details of individual interactions. In Study 2, we show that \ac{UX} experts and data scientists can effectively use ICEBOAT to visualize large amounts of automotive usage data to evaluate touchscreen-based \acp{IVIS}. Most importantly, ICEBOAT empowers \ac{UX} experts and contributes to the democratization of data in the automotive domain.

\begin{acks}
We want to thank Fabian Ober for his work on the user flow recording option.
\end{acks}

\bibliographystyle{ACM-Reference-Format}
\bibliography{autoui23}


\begin{thebibliography}{48}


\ifx \showCODEN    \undefined \def \showCODEN     #1{\unskip}     \fi
\ifx \showDOI      \undefined \def \showDOI       #1{#1}\fi
\ifx \showISBNx    \undefined \def \showISBNx     #1{\unskip}     \fi
\ifx \showISBNxiii \undefined \def \showISBNxiii  #1{\unskip}     \fi
\ifx \showISSN     \undefined \def \showISSN      #1{\unskip}     \fi
\ifx \showLCCN     \undefined \def \showLCCN      #1{\unskip}     \fi
\ifx \shownote     \undefined \def \shownote      #1{#1}          \fi
\ifx \showarticletitle \undefined \def \showarticletitle #1{#1}   \fi
\ifx \showURL      \undefined \def \showURL       {\relax}        \fi
\providecommand\bibfield[2]{#2}
\providecommand\bibinfo[2]{#2}
\providecommand\natexlab[1]{#1}
\providecommand\showeprint[2][]{arXiv:#2}

\bibitem[39(2020)]%
        {ISO15007}
\bibfield{author}{\bibinfo{person}{ISO/TC~22/SC 39}.}
  \bibinfo{year}{2020}\natexlab{}.
\newblock \bibinfo{booktitle}{\emph{{{ISO}} 15007:2020 {{Road}} Vehicles
  \textemdash{} {{Measurement}} and Analysis of Driver Visual Behaviour with
  Respect to Transport Information and Control Systems}}.
\newblock \bibinfo{type}{Standard}. \bibinfo{institution}{{International
  Organization for Standardization}}, \bibinfo{address}{{Geneva, CH}}.
\newblock


\bibitem[4(2018)]%
        {ISO9241-11.2018}
\bibfield{author}{\bibinfo{person}{ISO/TC~159/SC 4}.}
  \bibinfo{year}{2018}\natexlab{}.
\newblock \bibinfo{booktitle}{\emph{{{ISO}} 9241-11:2018 {{Ergonomics}} of
  Human-System Interaction \textemdash{} {{Part}} 11: {{Usability}}:
  {{Definitions}} and Concepts}}.
\newblock \bibinfo{type}{{T}echnical {R}eport}.
  \bibinfo{institution}{{International Organization for Standardization}}.
\newblock


\bibitem[Abras et~al\mbox{.}(2004)]%
        {abras.2004}
\bibfield{author}{\bibinfo{person}{Chadia Abras}, \bibinfo{person}{Diane
  {Maloney-Krichmar}}, {and} \bibinfo{person}{Jenny Preece}.}
  \bibinfo{year}{2004}\natexlab{}.
\newblock \showarticletitle{User-{{Centered Design}}}.
\newblock In \bibinfo{booktitle}{\emph{Encyclopedia of {{Human-Computer
  Interaction}}}}. \bibinfo{publisher}{{Sage Publications}},
  \bibinfo{address}{{Thousand Oaks, CA, USA}}.
\newblock


\bibitem[Administration(2014)]%
        {NHTSA.2014}
\bibfield{author}{\bibinfo{person}{National Highway Traffic~Safety
  Administration}.} \bibinfo{year}{2014}\natexlab{}.
\newblock \bibinfo{title}{Visual-{{Manual NHTSA Driver Distraction Guidelines}}
  for {{In-Vehicle Electronic Devices}}}.
\newblock
\newblock


\bibitem[Ahlemann et~al\mbox{.}(2021)]%
        {ahlemann.2021}
\bibfield{editor}{\bibinfo{person}{Frederik Ahlemann},
  \bibinfo{person}{Reinhard Sch{\"u}tte}, {and} \bibinfo{person}{Stefan
  Stieglitz}} (Eds.). \bibinfo{year}{2021}\natexlab{}.
\newblock \bibinfo{booktitle}{\emph{Innovation {{Through Information Systems}}:
  {{Volume III}}: {{A Collection}} of {{Latest Research}} on {{Management
  Issues}}}}. \bibinfo{series}{Lecture {{Notes}} in {{Information Systems}} and
  {{Organisation}}}, Vol.~\bibinfo{volume}{48}.
\newblock \bibinfo{publisher}{{Springer International Publishing}},
  \bibinfo{address}{{Cham}}.
\newblock
\showISBNx{978-3-030-86799-7 978-3-030-86800-0}
\urldef\tempurl%
\url{https://doi.org/10.1007/978-3-030-86800-0}
\showDOI{\tempurl}


\bibitem[Albers et~al\mbox{.}(2020)]%
        {albers.2020}
\bibfield{author}{\bibinfo{person}{Deike Albers}, \bibinfo{person}{Jonas
  Radlmayr}, \bibinfo{person}{Alexandra Loew}, \bibinfo{person}{Sebastian
  Hergeth}, \bibinfo{person}{Frederik Naujoks}, \bibinfo{person}{Andreas
  Keinath}, {and} \bibinfo{person}{Klaus Bengler}.}
  \bibinfo{year}{2020}\natexlab{}.
\newblock \showarticletitle{Usability {{Evaluation}} - {{Advances}} in
  {{Experimental Design}} in the {{Context}} of {{Automated Driving
  Human-Machine Interfaces}}}.
\newblock \bibinfo{journal}{\emph{Information}} \bibinfo{volume}{11},
  \bibinfo{number}{5} (\bibinfo{date}{April} \bibinfo{year}{2020}),
  \bibinfo{pages}{240}.
\newblock
\urldef\tempurl%
\url{https://doi.org/10.3390/info11050240}
\showDOI{\tempurl}


\bibitem[Arbesser et~al\mbox{.}(2017)]%
        {arbesser.2017}
\bibfield{author}{\bibinfo{person}{Clemens Arbesser}, \bibinfo{person}{Thomas
  M{\"u}hlbacher}, \bibinfo{person}{Stefan Komornyik}, {and}
  \bibinfo{person}{Harald Piringer}.} \bibinfo{year}{2017}\natexlab{}.
\newblock \showarticletitle{Visual Analytics for Domain Experts: {{Challenges}}
  and Lessons Learned}. In \bibinfo{booktitle}{\emph{Proceedings of the Second
  International Symposium on {{Virtual Reality}} \& {{Visual Computing}}}},
  \bibfield{editor}{\bibinfo{person}{VR~Kebao~(Tiajin) Science} {and}
  \bibinfo{person}{Ltd Technology~CO.}} (Eds.). \bibinfo{publisher}{{VR Kebao
  (Tiajin) Science and Technology CO.,Ltd}}, \bibinfo{address}{{}},
  \bibinfo{pages}{1--6}.
\newblock


\bibitem[Borsci et~al\mbox{.}(2015)]%
        {borsci.2015}
\bibfield{author}{\bibinfo{person}{Simone Borsci}, \bibinfo{person}{Stefano
  Federici}, \bibinfo{person}{Silvia Bacci}, \bibinfo{person}{Michela Gnaldi},
  {and} \bibinfo{person}{Francesco Bartolucci}.}
  \bibinfo{year}{2015}\natexlab{}.
\newblock \showarticletitle{Assessing {{User Satisfaction}} in the {{Era}} of
  {{User Experience}}: {{Comparison}} of the {{SUS}}, {{UMUX}}, and
  {{UMUX-LITE}} as a {{Function}} of {{Product Experience}}}.
\newblock \bibinfo{journal}{\emph{International Journal of Human-Computer
  Interaction}} \bibinfo{volume}{31}, \bibinfo{number}{8} (\bibinfo{date}{Aug.}
  \bibinfo{year}{2015}), \bibinfo{pages}{484--495}.
\newblock
\showISSN{1044-7318, 1532-7590}
\urldef\tempurl%
\url{https://doi.org/10.1080/10447318.2015.1064648}
\showDOI{\tempurl}


\bibitem[Brooke(1996)]%
        {brooke.1996}
\bibfield{author}{\bibinfo{person}{john Brooke}.}
  \bibinfo{year}{1996}\natexlab{}.
\newblock \showarticletitle{{{SUS}}: {{A}} '{{Quick}} and {{Dirty}}'
  {{Usability Scale}}}.
\newblock In \bibinfo{booktitle}{\emph{Usability {{Evaluation}} in
  {{Industry}}}}. \bibinfo{publisher}{{Taylor \& Francis}},
  \bibinfo{address}{{London, UK}}, \bibinfo{pages}{189--194}.
\newblock
\showISBNx{978-0-429-15701-1}


\bibitem[Broy et~al\mbox{.}(2007)]%
        {broy.2007}
\bibfield{author}{\bibinfo{person}{Manfred Broy}, \bibinfo{person}{Ingolf~H.
  Kruger}, \bibinfo{person}{Alexander Pretschner}, {and}
  \bibinfo{person}{Christian Salzmann}.} \bibinfo{year}{2007}\natexlab{}.
\newblock \showarticletitle{Engineering {{Automotive Software}}}.
\newblock \bibinfo{journal}{\emph{Proc. IEEE}} \bibinfo{volume}{95},
  \bibinfo{number}{2} (\bibinfo{date}{Feb.} \bibinfo{year}{2007}),
  \bibinfo{pages}{356--373}.
\newblock
\showISSN{0018-9219}
\urldef\tempurl%
\url{https://doi.org/10.1109/JPROC.2006.888386}
\showDOI{\tempurl}


\bibitem[Bryant and Wrigley(2014)]%
        {bryant.2014}
\bibfield{author}{\bibinfo{person}{Scott Bryant} {and} \bibinfo{person}{Cara
  Wrigley}.} \bibinfo{year}{2014}\natexlab{}.
\newblock \showarticletitle{Driving {{Toward User-Centered Engineering}} in
  {{Automotive Design}}}.
\newblock \bibinfo{journal}{\emph{Design Management Journal}}
  \bibinfo{volume}{9}, \bibinfo{number}{1} (\bibinfo{date}{Oct.}
  \bibinfo{year}{2014}), \bibinfo{pages}{74--84}.
\newblock
\showISSN{19425074}
\urldef\tempurl%
\url{https://doi.org/10.1111/dmj.12007}
\showDOI{\tempurl}


\bibitem[Cleland et~al\mbox{.}(2019)]%
        {cleland.2019}
\bibfield{author}{\bibinfo{person}{Brian Cleland}, \bibinfo{person}{Jonathan
  Wallace}, \bibinfo{person}{Raymond Bond}, \bibinfo{person}{Salla
  Muuraiskangas}, \bibinfo{person}{Juha Pajula}, \bibinfo{person}{Gorka
  Epelde}, \bibinfo{person}{M{\'o}nica Arr{\'u}e}, \bibinfo{person}{Roberto
  {\'A}lvarez}, \bibinfo{person}{Michaela Black}, \bibinfo{person}{Maurice~D.
  Mulvenna}, \bibinfo{person}{Deborah Rankin}, {and} \bibinfo{person}{Paul
  Carlin}.} \bibinfo{year}{2019}\natexlab{}.
\newblock \showarticletitle{Usability {{Evaluation}} of a {{Co-created Big Data
  Analytics Platform}} for {{Health Policy-Making}}}.
\newblock In \bibinfo{booktitle}{\emph{Human {{Interface}} and the
  {{Management}} of {{Information}}. {{Visual Information}} and {{Knowledge
  Management}}}}, \bibfield{editor}{\bibinfo{person}{Sakae Yamamoto} {and}
  \bibinfo{person}{Hirohiko Mori}} (Eds.). Vol.~\bibinfo{volume}{11569}.
  \bibinfo{publisher}{{Springer International Publishing}},
  \bibinfo{address}{{Cham}}, \bibinfo{pages}{194--207}.
\newblock
\showISBNx{978-3-030-22659-6 978-3-030-22660-2}
\urldef\tempurl%
\url{https://doi.org/10.1007/978-3-030-22660-2_13}
\showDOI{\tempurl}


\bibitem[Crosslin et~al\mbox{.}(2021)]%
        {crosslin.2021}
\bibfield{author}{\bibinfo{person}{Matt Crosslin}, \bibinfo{person}{Kimberly
  Breuer}, \bibinfo{person}{Nikola Miliki{\'c}}, {and}
  \bibinfo{person}{Justin~T. Dellinger}.} \bibinfo{year}{2021}\natexlab{}.
\newblock \showarticletitle{Understanding Student Learning Pathways in
  Traditional Online History Courses: Utilizing Process Mining Analysis on
  Clickstream Data}.
\newblock \bibinfo{journal}{\emph{Journal of Research in Innovative Teaching \&
  Learning}} \bibinfo{volume}{14}, \bibinfo{number}{3} (\bibinfo{date}{Nov.}
  \bibinfo{year}{2021}), \bibinfo{pages}{399--414}.
\newblock
\showISSN{2397-7604}
\urldef\tempurl%
\url{https://doi.org/10.1108/JRIT-03-2021-0024}
\showDOI{\tempurl}


\bibitem[Cui(2019)]%
        {cui.2019}
\bibfield{author}{\bibinfo{person}{Wenqiang Cui}.}
  \bibinfo{year}{2019}\natexlab{}.
\newblock \showarticletitle{Visual {{Analytics}}: {{A Comprehensive
  Overview}}}.
\newblock \bibinfo{journal}{\emph{IEEE Access}}  \bibinfo{volume}{7}
  (\bibinfo{year}{2019}), \bibinfo{pages}{81555--81573}.
\newblock
\showISSN{2169-3536}
\urldef\tempurl%
\url{https://doi.org/10.1109/ACCESS.2019.2923736}
\showDOI{\tempurl}


\bibitem[Dremel et~al\mbox{.}(2017)]%
        {dremel.2017}
\bibfield{author}{\bibinfo{person}{Christian Dremel}, \bibinfo{person}{Jochen
  Wulf}, \bibinfo{person}{Matthias~M Herterich}, \bibinfo{person}{Jean-Claude
  Waizmann}, {and} \bibinfo{person}{Walter Brenner}.}
  \bibinfo{year}{2017}\natexlab{}.
\newblock \showarticletitle{How {{AUDI AG}} Established Big Data Analytics in
  Its Digital Transformation.}
\newblock \bibinfo{journal}{\emph{MIS Quarterly Executive}}
  \bibinfo{volume}{16}, \bibinfo{number}{2} (\bibinfo{year}{2017}),
  \bibinfo{pages}{81--101}.
\newblock


\bibitem[Ebel et~al\mbox{.}(2020)]%
        {ebel.2020}
\bibfield{author}{\bibinfo{person}{Patrick Ebel}, \bibinfo{person}{Florian
  Brokhausen}, {and} \bibinfo{person}{Andreas Vogelsang}.}
  \bibinfo{year}{2020}\natexlab{}.
\newblock \showarticletitle{The {{Role}} and {{Potentials}} of {{Field User
  Interaction Data}} in the {{Automotive UX Development Lifecycle}}: {{An
  Industry Perspective}}}. In \bibinfo{booktitle}{\emph{12th {{International
  Conference}} on {{Automotive User Interfaces}} and {{Interactive Vehicular
  Applications}}}}. \bibinfo{publisher}{{ACM}}, \bibinfo{address}{{Virtual
  Event DC USA}}, \bibinfo{pages}{141--150}.
\newblock
\showISBNx{978-1-4503-8065-2}
\urldef\tempurl%
\url{https://doi.org/10.1145/3409120.3410638}
\showDOI{\tempurl}


\bibitem[Ebel et~al\mbox{.}(2022)]%
        {ebel.ICEBOAT.2022}
\bibfield{author}{\bibinfo{person}{Patrick Ebel}, \bibinfo{person}{Kim~Julian
  G\"{u}lle}, \bibinfo{person}{Christoph Lingenfelder}, {and}
  \bibinfo{person}{Andreas Vogelsang}.} \bibinfo{year}{2022}\natexlab{}.
\newblock \showarticletitle{ICEBOAT: An Interactive User Behavior Analysis Tool
  for Automotive User Interfaces}. In \bibinfo{booktitle}{\emph{Adjunct
  Proceedings of the 35th Annual ACM Symposium on User Interface Software and
  Technology}} (Bend, OR, USA) \emph{(\bibinfo{series}{UIST '22 Adjunct})}.
  \bibinfo{publisher}{Association for Computing Machinery},
  \bibinfo{address}{New York, NY, USA}, Article \bibinfo{articleno}{50},
  \bibinfo{numpages}{3}~pages.
\newblock
\showISBNx{9781450393218}
\urldef\tempurl%
\url{https://doi.org/10.1145/3526114.3558739}
\showDOI{\tempurl}


\bibitem[Ebel et~al\mbox{.}(2021a)]%
        {ebel.2021}
\bibfield{author}{\bibinfo{person}{Patrick Ebel}, \bibinfo{person}{Christoph
  Lingenfelder}, {and} \bibinfo{person}{Andreas Vogelsang}.}
  \bibinfo{year}{2021}\natexlab{a}.
\newblock \showarticletitle{Visualizing {{Event Sequence Data}} for {{User
  Behavior Evaluation}} of {{In-Vehicle Information Systems}}}. In
  \bibinfo{booktitle}{\emph{13th {{International Conference}} on {{Automotive
  User Interfaces}} and {{Interactive Vehicular Applications}}}}.
  \bibinfo{publisher}{{ACM}}, \bibinfo{address}{{Leeds United Kingdom}},
  \bibinfo{pages}{219--229}.
\newblock
\showISBNx{978-1-4503-8063-8}
\urldef\tempurl%
\url{https://doi.org/10.1145/3409118.3475140}
\showDOI{\tempurl}


\bibitem[Ebel et~al\mbox{.}(2023)]%
        {ebel.2023}
\bibfield{author}{\bibinfo{person}{Patrick Ebel}, \bibinfo{person}{Christoph
  Lingenfelder}, {and} \bibinfo{person}{Andreas Vogelsang}.}
  \bibinfo{year}{2023}\natexlab{}.
\newblock \showarticletitle{Multitasking While Driving: How Drivers
  Self-Regulate Their Interaction with In-Vehicle Touchscreens in Automated
  Driving}.
\newblock \bibinfo{journal}{\emph{International Journal of Human–Computer
  Interaction}} \bibinfo{volume}{0}, \bibinfo{number}{0}
  (\bibinfo{year}{2023}), \bibinfo{pages}{1--18}.
\newblock
\urldef\tempurl%
\url{https://doi.org/10.1080/10447318.2023.2215634}
\showDOI{\tempurl}
\showeprint{https://doi.org/10.1080/10447318.2023.2215634}


\bibitem[Ebel et~al\mbox{.}(2021b)]%
        {ebel.2021a}
\bibfield{author}{\bibinfo{person}{Patrick Ebel}, \bibinfo{person}{Julia
  Orlovska}, \bibinfo{person}{Sebastian H{\"u}nemeyer}, \bibinfo{person}{Casper
  Wickman}, \bibinfo{person}{Andreas Vogelsang}, {and} \bibinfo{person}{Rikard
  S{\"o}derberg}.} \bibinfo{year}{2021}\natexlab{b}.
\newblock \showarticletitle{Automotive {{UX}} Design and Data-Driven
  Development: {{Narrowing}} the Gap to Support Practitioners}.
\newblock \bibinfo{journal}{\emph{Transportation Research Interdisciplinary
  Perspectives}}  \bibinfo{volume}{11} (\bibinfo{date}{Sept.}
  \bibinfo{year}{2021}), \bibinfo{pages}{100455}.
\newblock
\showISSN{25901982}
\urldef\tempurl%
\url{https://doi.org/10.1016/j.trip.2021.100455}
\showDOI{\tempurl}


\bibitem[Fisher et~al\mbox{.}(2012)]%
        {fisher.2012}
\bibfield{author}{\bibinfo{person}{Danyel Fisher}, \bibinfo{person}{Rob
  DeLine}, \bibinfo{person}{Mary Czerwinski}, {and} \bibinfo{person}{Steven
  Drucker}.} \bibinfo{year}{2012}\natexlab{}.
\newblock \showarticletitle{Interactions with Big Data Analytics}.
\newblock \bibinfo{journal}{\emph{Interactions}} \bibinfo{volume}{19},
  \bibinfo{number}{3} (\bibinfo{date}{May} \bibinfo{year}{2012}),
  \bibinfo{pages}{50--59}.
\newblock
\showISSN{1072-5520, 1558-3449}
\urldef\tempurl%
\url{https://doi.org/10.1145/2168931.2168943}
\showDOI{\tempurl}


\bibitem[Frison et~al\mbox{.}(2019)]%
        {frison.2019}
\bibfield{author}{\bibinfo{person}{Anna-Katharina Frison},
  \bibinfo{person}{Philipp Wintersberger}, \bibinfo{person}{Andreas Riener},
  \bibinfo{person}{Clemens Schartm{\"u}ller}, \bibinfo{person}{Linda~Ng Boyle},
  \bibinfo{person}{Erika Miller}, {and} \bibinfo{person}{Klemens Weigl}.}
  \bibinfo{year}{2019}\natexlab{}.
\newblock \showarticletitle{In {{UX We Trust}}: {{Investigation}} of
  {{Aesthetics}} and {{Usability}} of {{Driver-Vehicle Interfaces}} and {{Their
  Impact}} on the {{Perception}} of {{Automated Driving}}}. In
  \bibinfo{booktitle}{\emph{Proceedings of the 2019 {{CHI Conference}} on
  {{Human Factors}} in {{Computing Systems}}}}. \bibinfo{publisher}{{ACM}},
  \bibinfo{address}{{Glasgow Scotland Uk}}, \bibinfo{pages}{1--13}.
\newblock
\showISBNx{978-1-4503-5970-2}
\urldef\tempurl%
\url{https://doi.org/10.1145/3290605.3300374}
\showDOI{\tempurl}


\bibitem[Grahn and Kujala(2020)]%
        {grahn.2020}
\bibfield{author}{\bibinfo{person}{Hilkka Grahn} {and} \bibinfo{person}{Tuomo
  Kujala}.} \bibinfo{year}{2020}\natexlab{}.
\newblock \showarticletitle{Impacts of Touch Screen Size, User Interface
  Design, and Subtask Boundaries on in-Car Tasks Visual Demand and Driver
  Distraction}.
\newblock \bibinfo{journal}{\emph{International Journal of Human-Computer
  Studies}}  \bibinfo{volume}{142} (\bibinfo{date}{Oct.} \bibinfo{year}{2020}),
  \bibinfo{pages}{102467}.
\newblock
\urldef\tempurl%
\url{https://doi.org/10.1016/j.ijhcs.2020.102467}
\showDOI{\tempurl}


\bibitem[Guo et~al\mbox{.}(2022)]%
        {guo.2022}
\bibfield{author}{\bibinfo{person}{Yi Guo}, \bibinfo{person}{Shunan Guo},
  \bibinfo{person}{Zhuochen Jin}, \bibinfo{person}{Smiti Kaul},
  \bibinfo{person}{David Gotz}, {and} \bibinfo{person}{Nan Cao}.}
  \bibinfo{year}{2022}\natexlab{}.
\newblock \showarticletitle{Survey on {{Visual Analysis}} of {{Event Sequence
  Data}}}.
\newblock \bibinfo{journal}{\emph{IEEE Transactions on Visualization and
  Computer Graphics}} \bibinfo{volume}{28}, \bibinfo{number}{12}
  (\bibinfo{date}{Dec.} \bibinfo{year}{2022}), \bibinfo{pages}{5091--5112}.
\newblock
\showISSN{1077-2626, 1941-0506, 2160-9306}
\urldef\tempurl%
\url{https://doi.org/10.1109/TVCG.2021.3100413}
\showDOI{\tempurl}


\bibitem[Harvey and Stanton(2016)]%
        {harvey.2016}
\bibfield{author}{\bibinfo{person}{Catherine Harvey} {and}
  \bibinfo{person}{Neville~A. Stanton}.} \bibinfo{year}{2016}\natexlab{}.
\newblock \bibinfo{booktitle}{\emph{Usability {{Evaluation}} for {{In-Vehicle
  Systems}}} (\bibinfo{edition}{zeroth} ed.)}.
\newblock \bibinfo{publisher}{{CRC Press}}, \bibinfo{address}{{}}.
\newblock
\showISBNx{978-0-429-09898-7}
\urldef\tempurl%
\url{https://doi.org/10.1201/b14644}
\showDOI{\tempurl}


\bibitem[Harvey et~al\mbox{.}(2011)]%
        {harvey.2011}
\bibfield{author}{\bibinfo{person}{Catherine Harvey},
  \bibinfo{person}{Neville~A. Stanton}, \bibinfo{person}{Carl~A. Pickering},
  \bibinfo{person}{Mike McDonald}, {and} \bibinfo{person}{Pengjun Zheng}.}
  \bibinfo{year}{2011}\natexlab{}.
\newblock \showarticletitle{In-{{Vehicle Information Systems}} to {{Meet}} the
  {{Needs}} of {{Drivers}}}.
\newblock \bibinfo{journal}{\emph{International Journal of Human-Computer
  Interaction}} \bibinfo{volume}{27}, \bibinfo{number}{6} (\bibinfo{date}{June}
  \bibinfo{year}{2011}), \bibinfo{pages}{505--522}.
\newblock
\showISSN{1044-7318, 1532-7590}
\urldef\tempurl%
\url{https://doi.org/10.1080/10447318.2011.555296}
\showDOI{\tempurl}


\bibitem[Jansen et~al\mbox{.}(2023)]%
        {jansen.2023}
\bibfield{author}{\bibinfo{person}{Pascal Jansen}, \bibinfo{person}{Julian
  Britten}, \bibinfo{person}{Alexander H\"{a}usele}, \bibinfo{person}{Thilo
  Segschneider}, \bibinfo{person}{Mark Colley}, {and} \bibinfo{person}{Enrico
  Rukzio}.} \bibinfo{year}{2023}\natexlab{}.
\newblock \showarticletitle{AutoVis: Enabling Mixed-Immersive Analysis of
  Automotive User Interface Interaction Studies}. In
  \bibinfo{booktitle}{\emph{Proceedings of the 2023 CHI Conference on Human
  Factors in Computing Systems}} (Hamburg, Germany) \emph{(\bibinfo{series}{CHI
  '23})}. \bibinfo{publisher}{Association for Computing Machinery},
  \bibinfo{address}{New York, NY, USA}, Article \bibinfo{articleno}{378},
  \bibinfo{numpages}{23}~pages.
\newblock
\showISBNx{9781450394215}
\urldef\tempurl%
\url{https://doi.org/10.1145/3544548.3580760}
\showDOI{\tempurl}


\bibitem[Keim et~al\mbox{.}(2008)]%
        {keim.2008}
\bibfield{author}{\bibinfo{person}{Daniel Keim}, \bibinfo{person}{Gennady
  Andrienko}, \bibinfo{person}{Jean-Daniel Fekete}, \bibinfo{person}{Carsten
  G{\"o}rg}, \bibinfo{person}{J{\"o}rn Kohlhammer}, {and} \bibinfo{person}{Guy
  Melan{\c{c}}on}.} \bibinfo{year}{2008}\natexlab{}.
\newblock \showarticletitle{Visual Analytics: Definition, Process, and
  Challenges}.
\newblock In \bibinfo{booktitle}{\emph{Information Visualization:
  Human-Centered Issues and Perspectives}},
  \bibfield{editor}{\bibinfo{person}{Andreas Kerren}, \bibinfo{person}{John~T.
  Stasko}, \bibinfo{person}{Jean-Daniel Fekete}, {and} \bibinfo{person}{Chris
  North}} (Eds.). \bibinfo{publisher}{Springer Berlin Heidelberg},
  \bibinfo{address}{Berlin, Heidelberg}, \bibinfo{pages}{154--175}.
\newblock
\showISBNx{978-3-540-70956-5}
\urldef\tempurl%
\url{https://doi.org/10.1007/978-3-540-70956-5_7}
\showDOI{\tempurl}


\bibitem[Klauer et~al\mbox{.}(2006)]%
        {Klauer.2006}
\bibfield{author}{\bibinfo{person}{Sheila Klauer}, \bibinfo{person}{Thomas
  Dingus}, \bibinfo{person}{T Neale}, \bibinfo{person}{J. Sudweeks}, {and}
  \bibinfo{person}{D Ramsey}.} \bibinfo{year}{2006}\natexlab{}.
\newblock \bibinfo{booktitle}{\emph{The Impact of Driver Inattention on
  Near-Crash/Crash Risk: {{An}} Analysis Using the 100-{{Car}} Naturalistic
  Driving Study Data}}.
\newblock \bibinfo{type}{{T}echnical {R}eport}. \bibinfo{institution}{{U.S.
  Department of Transportation, National Highway Traffic Safety Administration
  / Virginia Tech Transportation Institute}}, \bibinfo{address}{{3500
  Transportation Research Plaza (0536) Blacksburg, Virginia 24061}}.
\newblock


\bibitem[Lewis and Sauro(2018)]%
        {lewis.2018}
\bibfield{author}{\bibinfo{person}{James~R Lewis} {and} \bibinfo{person}{Jeff
  Sauro}.} \bibinfo{year}{2018}\natexlab{}.
\newblock \showarticletitle{Item {{Benchmarks}} for the {{System Usability
  Scale}}}.
\newblock \bibinfo{journal}{\emph{Journal of Usability Studies}}
  \bibinfo{volume}{13}, \bibinfo{number}{3} (\bibinfo{date}{May}
  \bibinfo{year}{2018}), \bibinfo{pages}{158--167}.
\newblock


\bibitem[Liu et~al\mbox{.}(2017)]%
        {liu.2017}
\bibfield{author}{\bibinfo{person}{Zhicheng Liu}, \bibinfo{person}{Yang Wang},
  \bibinfo{person}{Mira Dontcheva}, \bibinfo{person}{Matthew Hoffman},
  \bibinfo{person}{Seth Walker}, {and} \bibinfo{person}{Alan Wilson}.}
  \bibinfo{year}{2017}\natexlab{}.
\newblock \showarticletitle{Patterns and {{Sequences}}: {{Interactive
  Exploration}} of {{Clickstreams}} to {{Understand Common Visitor Paths}}}.
\newblock \bibinfo{journal}{\emph{IEEE Transactions on Visualization and
  Computer Graphics}} \bibinfo{volume}{23}, \bibinfo{number}{1}
  (\bibinfo{date}{Jan.} \bibinfo{year}{2017}), \bibinfo{pages}{321--330}.
\newblock
\showISSN{1077-2626}
\urldef\tempurl%
\url{https://doi.org/10.1109/TVCG.2016.2598797}
\showDOI{\tempurl}


\bibitem[Matejka et~al\mbox{.}(2013)]%
        {matejka.2013}
\bibfield{author}{\bibinfo{person}{Justin Matejka}, \bibinfo{person}{Tovi
  Grossman}, {and} \bibinfo{person}{George Fitzmaurice}.}
  \bibinfo{year}{2013}\natexlab{}.
\newblock \showarticletitle{Patina: Dynamic Heatmaps for Visualizing
  Application Usage}. In \bibinfo{booktitle}{\emph{Proceedings of the {{SIGCHI
  Conference}} on {{Human Factors}} in {{Computing Systems}}}}.
  \bibinfo{publisher}{{ACM}}, \bibinfo{address}{{Paris France}},
  \bibinfo{pages}{3227--3236}.
\newblock
\showISBNx{978-1-4503-1899-0}
\urldef\tempurl%
\url{https://doi.org/10.1145/2470654.2466442}
\showDOI{\tempurl}


\bibitem[Mattos et~al\mbox{.}(2017)]%
        {mattos.2017}
\bibfield{author}{\bibinfo{person}{David~Issa Mattos}, \bibinfo{person}{Jan
  Bosch}, {and} \bibinfo{person}{Helena~Holmstr{\"o}m Olsson}.}
  \bibinfo{year}{2017}\natexlab{}.
\newblock \showarticletitle{Your {{System Gets Better Every Day You Use It}}:
  {{Towards Automated Continuous Experimentation}}}. In
  \bibinfo{booktitle}{\emph{2017 43rd {{Euromicro Conference}} on {{Software
  Engineering}} and {{Advanced Applications}} ({{SEAA}})}}.
  \bibinfo{publisher}{{IEEE}}, \bibinfo{address}{Vienna, Austria},
  \bibinfo{pages}{256--265}.
\newblock
\urldef\tempurl%
\url{https://doi.org/10.1109/SEAA.2017.15}
\showDOI{\tempurl}


\bibitem[McLellan et~al\mbox{.}(2012)]%
        {mclellan.2012}
\bibfield{author}{\bibinfo{person}{Sam McLellan}, \bibinfo{person}{Andrew
  Muddimer}, {and} \bibinfo{person}{S~Camille Peres}.}
  \bibinfo{year}{2012}\natexlab{}.
\newblock \showarticletitle{The {{Effect}} of {{Experience}} on {{System
  Usability Scale Ratings}}}.
\newblock \bibinfo{journal}{\emph{Journal of Usability Studies}}
  \bibinfo{volume}{7}, \bibinfo{number}{2} (\bibinfo{date}{Feb.}
  \bibinfo{year}{2012}), \bibinfo{pages}{56--67}.
\newblock


\bibitem[{On-Road Automated Driving (ORAD) committee}(2021)]%
        {SAE.2021}
\bibfield{author}{\bibinfo{person}{{On-Road Automated Driving (ORAD)
  committee}}.} \bibinfo{year}{2021}\natexlab{}.
\newblock \bibinfo{booktitle}{\emph{Taxonomy and {{Definitions}} for {{Terms
  Related}} to {{Driving Automation Systems}} for {{On-Road Motor Vehicles}}}}.
\newblock \bibinfo{type}{{T}echnical {R}eport}. \bibinfo{institution}{{SAE
  International}}.
\newblock
\urldef\tempurl%
\url{https://doi.org/10.4271/J3016_202104}
\showDOI{\tempurl}


\bibitem[Power(2020)]%
        {JDPower.2020}
\bibfield{author}{\bibinfo{person}{J.D. Power}.}
  \bibinfo{year}{2020}\natexlab{}.
\newblock \bibinfo{booktitle}{\emph{2020 {{U}}.{{S}}. {{Initial Quality
  Study}}}}.
\newblock \bibinfo{type}{{T}echnical {R}eport}. \bibinfo{institution}{{J.D.
  Power}}.
\newblock


\bibitem[Power(2021)]%
        {JDPower.2021}
\bibfield{author}{\bibinfo{person}{J.D. Power}.}
  \bibinfo{year}{2021}\natexlab{}.
\newblock \bibinfo{booktitle}{\emph{2021 {{U}}.{{S}}. {{Initial Quality
  Study}}}}.
\newblock \bibinfo{type}{{T}echnical {R}eport}. \bibinfo{institution}{{J.D.
  Power}}.
\newblock


\bibitem[Power(2022)]%
        {JDPower.2022}
\bibfield{author}{\bibinfo{person}{J.D. Power}.}
  \bibinfo{year}{2022}\natexlab{}.
\newblock \bibinfo{booktitle}{\emph{2022 {{U}}.{{S}}. {{Initial Quality
  Study}}}}.
\newblock \bibinfo{type}{{T}echnical {R}eport}. \bibinfo{institution}{{J.D.
  Power}}.
\newblock


\bibitem[Regan and {Oviedo-Trespalacios}(2022)]%
        {regan.2022}
\bibfield{author}{\bibinfo{person}{Michael~A. Regan} {and}
  \bibinfo{person}{Oscar {Oviedo-Trespalacios}}.}
  \bibinfo{year}{2022}\natexlab{}.
\newblock \showarticletitle{Driver {{Distraction}}: {{Mechanisms}},
  {{Evidence}}, {{Prevention}}, and {{Mitigation}}}.
\newblock In \bibinfo{booktitle}{\emph{The {{Vision Zero Handbook}}}},
  \bibfield{editor}{\bibinfo{person}{Karin Edvardsson~Bj{\"o}rnberg},
  \bibinfo{person}{Matts-{\AA}ke Belin}, \bibinfo{person}{Sven~Ove Hansson},
  {and} \bibinfo{person}{Claes Tingvall}} (Eds.). \bibinfo{publisher}{{Springer
  International Publishing}}, \bibinfo{address}{{Cham}},
  \bibinfo{pages}{1--62}.
\newblock
\showISBNx{978-3-030-23176-7}
\urldef\tempurl%
\url{https://doi.org/10.1007/978-3-030-23176-7_38-1}
\showDOI{\tempurl}


\bibitem[Renner and Jacob(2020)]%
        {renner.2020}
\bibfield{author}{\bibinfo{person}{Karl-Heinz Renner} {and}
  \bibinfo{person}{Nora-Corina Jacob}.} \bibinfo{year}{2020}\natexlab{}.
\newblock \bibinfo{booktitle}{\emph{{Das Interview: Grundlagen und Anwendung in
  Psychologie und Sozialwissenschaften}}}.
\newblock \bibinfo{publisher}{{Springer Berlin Heidelberg}},
  \bibinfo{address}{{Berlin, Heidelberg}}.
\newblock
\showISBNx{978-3-662-60440-3 978-3-662-60441-0}
\urldef\tempurl%
\url{https://doi.org/10.1007/978-3-662-60441-0}
\showDOI{\tempurl}


\bibitem[Sadeghi et~al\mbox{.}(2023)]%
        {sadeghi.2023}
\bibfield{author}{\bibinfo{person}{Mersedeh Sadeghi}, \bibinfo{person}{Alessio
  Carenini}, \bibinfo{person}{Oscar Corcho}, \bibinfo{person}{Matteo Rossi},
  \bibinfo{person}{Riccardo Santoro}, {and} \bibinfo{person}{Andreas
  Vogelsang}.} \bibinfo{year}{2023}\natexlab{}.
\newblock \showarticletitle{Interoperability of Heterogeneous Systems of
  Systems: Review of Challenges, Emerging Requirements and Options}. In
  \bibinfo{booktitle}{\emph{Proceedings of the 38th ACM/SIGAPP Symposium on
  Applied Computing}} (Tallinn, Estonia) \emph{(\bibinfo{series}{SAC '23})}.
  \bibinfo{publisher}{Association for Computing Machinery},
  \bibinfo{address}{New York, NY, USA}, \bibinfo{pages}{741–750}.
\newblock
\showISBNx{9781450395175}
\urldef\tempurl%
\url{https://doi.org/10.1145/3555776.3577692}
\showDOI{\tempurl}


\bibitem[Saura(2021)]%
        {saura.2021}
\bibfield{author}{\bibinfo{person}{Jose~Ramon Saura}.}
  \bibinfo{year}{2021}\natexlab{}.
\newblock \showarticletitle{Using {{Data Sciences}} in {{Digital Marketing}}:
  {{Framework}}, Methods, and Performance Metrics}.
\newblock \bibinfo{journal}{\emph{Journal of Innovation \& Knowledge}}
  \bibinfo{volume}{6}, \bibinfo{number}{2} (\bibinfo{date}{April}
  \bibinfo{year}{2021}), \bibinfo{pages}{92--102}.
\newblock
\showISSN{2444569X}
\urldef\tempurl%
\url{https://doi.org/10.1016/j.jik.2020.08.001}
\showDOI{\tempurl}


\bibitem[Saura et~al\mbox{.}(2021)]%
        {saura.2021a}
\bibfield{author}{\bibinfo{person}{Jose~Ramon Saura}, \bibinfo{person}{Domingo
  {Ribeiro-Soriano}}, {and} \bibinfo{person}{Daniel {Palacios-Marqu{\'e}s}}.}
  \bibinfo{year}{2021}\natexlab{}.
\newblock \showarticletitle{From User-Generated Data to Data-Driven Innovation:
  {{A}} Research Agenda to Understand User Privacy in Digital Markets}.
\newblock \bibinfo{journal}{\emph{International Journal of Information
  Management}}  \bibinfo{volume}{60} (\bibinfo{date}{Oct.}
  \bibinfo{year}{2021}), \bibinfo{pages}{102331}.
\newblock
\showISSN{02684012}
\urldef\tempurl%
\url{https://doi.org/10.1016/j.ijinfomgt.2021.102331}
\showDOI{\tempurl}


\bibitem[Schermann et~al\mbox{.}(2018)]%
        {schermann.2018}
\bibfield{author}{\bibinfo{person}{Gerald Schermann},
  \bibinfo{person}{J{\"u}rgen Cito}, {and} \bibinfo{person}{Philipp Leitner}.}
  \bibinfo{year}{2018}\natexlab{}.
\newblock \showarticletitle{Continuous {{Experimentation}}: {{Challenges}},
  {{Implementation Techniques}}, and {{Current Research}}}.
\newblock \bibinfo{journal}{\emph{IEEE Software}} \bibinfo{volume}{35},
  \bibinfo{number}{2} (\bibinfo{date}{March} \bibinfo{year}{2018}),
  \bibinfo{pages}{26--31}.
\newblock
\showISSN{1937-4194}
\urldef\tempurl%
\url{https://doi.org/10.1109/MS.2018.111094748}
\showDOI{\tempurl}


\bibitem[Sedlmair et~al\mbox{.}(2011)]%
        {sedlmair.2011}
\bibfield{author}{\bibinfo{person}{Michael Sedlmair}, \bibinfo{person}{Petra
  Isenberg}, \bibinfo{person}{Dominikus Baur}, \bibinfo{person}{Michael
  Mauerer}, \bibinfo{person}{Christian Pigorsch}, {and}
  \bibinfo{person}{Andreas Butz}.} \bibinfo{year}{2011}\natexlab{}.
\newblock \showarticletitle{Cardiogram: Visual Analytics for Automotive
  Engineers}. In \bibinfo{booktitle}{\emph{Proceedings of the {{SIGCHI
  Conference}} on {{Human Factors}} in {{Computing Systems}}}}.
  \bibinfo{publisher}{{ACM}}, \bibinfo{address}{{Vancouver BC Canada}},
  \bibinfo{pages}{1727--1736}.
\newblock
\showISBNx{978-1-4503-0228-9}
\urldef\tempurl%
\url{https://doi.org/10.1145/1978942.1979194}
\showDOI{\tempurl}


\bibitem[Still and Crane(2017)]%
        {still.2017}
\bibfield{author}{\bibinfo{person}{Brian Still} {and} \bibinfo{person}{Kate
  Crane}.} \bibinfo{year}{2017}\natexlab{}.
\newblock \bibinfo{booktitle}{\emph{Fundamentals of {{User-Centered Design}}:
  {{A Practical Approach}}} (\bibinfo{edition}{first} ed.)}.
\newblock \bibinfo{publisher}{{CRC Press}}, \bibinfo{address}{{Boca Raton}}.
\newblock
\showISBNx{978-1-315-20092-7}
\urldef\tempurl%
\url{https://doi.org/10.4324/9781315200927}
\showDOI{\tempurl}


\bibitem[Wongsuphasawat and Gotz(2012)]%
        {wongsuphasawat.2012}
\bibfield{author}{\bibinfo{person}{K. Wongsuphasawat} {and} \bibinfo{person}{D.
  Gotz}.} \bibinfo{year}{2012}\natexlab{}.
\newblock \showarticletitle{Exploring {{Flow}}, {{Factors}}, and {{Outcomes}}
  of {{Temporal Event Sequences}} with the {{Outflow Visualization}}}.
\newblock \bibinfo{journal}{\emph{IEEE Transactions on Visualization and
  Computer Graphics}} \bibinfo{volume}{18}, \bibinfo{number}{12}
  (\bibinfo{date}{Dec.} \bibinfo{year}{2012}), \bibinfo{pages}{2659--2668}.
\newblock
\showISSN{1077-2626}
\urldef\tempurl%
\url{https://doi.org/10.1109/TVCG.2012.225}
\showDOI{\tempurl}


\bibitem[Zhang et~al\mbox{.}(2012)]%
        {zhang.2012}
\bibfield{author}{\bibinfo{person}{Leishi Zhang}, \bibinfo{person}{Andreas
  Stoffel}, \bibinfo{person}{Michael Behrisch}, \bibinfo{person}{Sebastian
  Mittelstadt}, \bibinfo{person}{Tobias Schreck}, \bibinfo{person}{Rene Pompl},
  \bibinfo{person}{Stefan Weber}, \bibinfo{person}{Holger Last}, {and}
  \bibinfo{person}{Daniel Keim}.} \bibinfo{year}{2012}\natexlab{}.
\newblock \showarticletitle{Visual Analytics for the Big Data Era \&\#x2014;
  {{A}} Comparative Review of State-of-the-Art Commercial Systems}. In
  \bibinfo{booktitle}{\emph{2012 {{IEEE Conference}} on {{Visual Analytics
  Science}} and {{Technology}} ({{VAST}})}}. \bibinfo{publisher}{{IEEE}},
  \bibinfo{address}{{Seattle, WA, USA}}, \bibinfo{pages}{173--182}.
\newblock
\showISBNx{978-1-4673-4753-2 978-1-4673-4752-5}
\urldef\tempurl%
\url{https://doi.org/10.1109/VAST.2012.6400554}
\showDOI{\tempurl}


\end{thebibliography}


\appendix
\section{Appendix}\label{ch:Appendix}

\subsection{Evaluation Tasks}\label{app:Tasks}

\begin{enumerate}
    \item Use the record button of the \ac{IVIS} emulator to record a flow beginning at the navigation system's start screen and ending with the "Let's Go" button.
    \item Use the record file to define and analyze the customer journey.
    \item What are the top 5 flows (by share)? Use the filters to only display these flows.
    \item Identify one bottleneck in the flows. Could you explain the potential causes?
    \item Compare the glance behavior (count) of the first 3 flows.
    \item Which sequence in flow 1 has the highest glance count? (open the Sequence Detail View).
    \item Identify one long glance and explain the driving situation. Are there possibly distracting interactions?
\end{enumerate}

\subsection{Context of Use Questionnaire}\label{app:ContextofUse}

\begin{table*}[!htb]
    \label{tab:context_of_use_questionnaire}
    \caption{The Context of Use Questionnaire}
    \centering
    \begin{tabular}{lp{.75\linewidth}}
        \toprule
        \textbf{No.} & \textbf{Question} \\
        \midrule
        1 & I think the system allows me to analyze telematics data on my own (independent from another person or department)\\
        2 & I think the system makes telematics data accessible\\
        \multicolumn{2}{l}{The system provides insights into telematics data\dots}\\
         3 & \dots which are new to me\\
         4 & \dots that help me to better understand how our customers interact with the infotainment system\\
        5 & \dots which help me to observe how our customers interact with the infotainment system in different driving situations\\
        6 & \dots which allow me to base my decisions on data\\
        7 & \dots which help me resolve discussions about feature priorities\\
        8 & The system helps me to identify usability issues in our infotainment system\\
        9 & Having the system available would accelerate our current workflow with telematics data analysis\\
        10 & In which phase of the design process would you use the system?\\    
        \bottomrule
    \end{tabular}
\end{table*}

\pagebreak

\subsection{Storyboard}\label{app:Storyboard}

	\begin{figure*}[htb!]
		\centering
		\includegraphics[width=\linewidth]{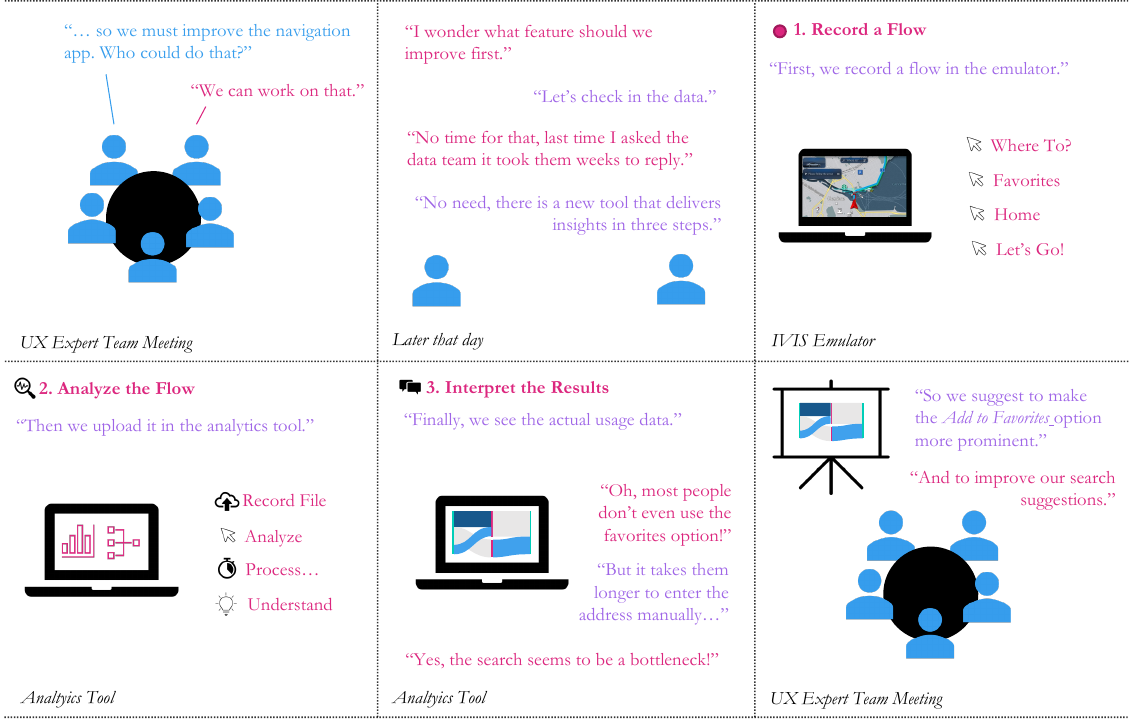}
		\caption{Storyboard}
        \label{fig:StoryBoard}
	\end{figure*}

\subsection{Sequence Diagram}\label{app:SequenceDiagram}

	\begin{figure*}[htb!]
		\centering
		\includegraphics[width=\linewidth]{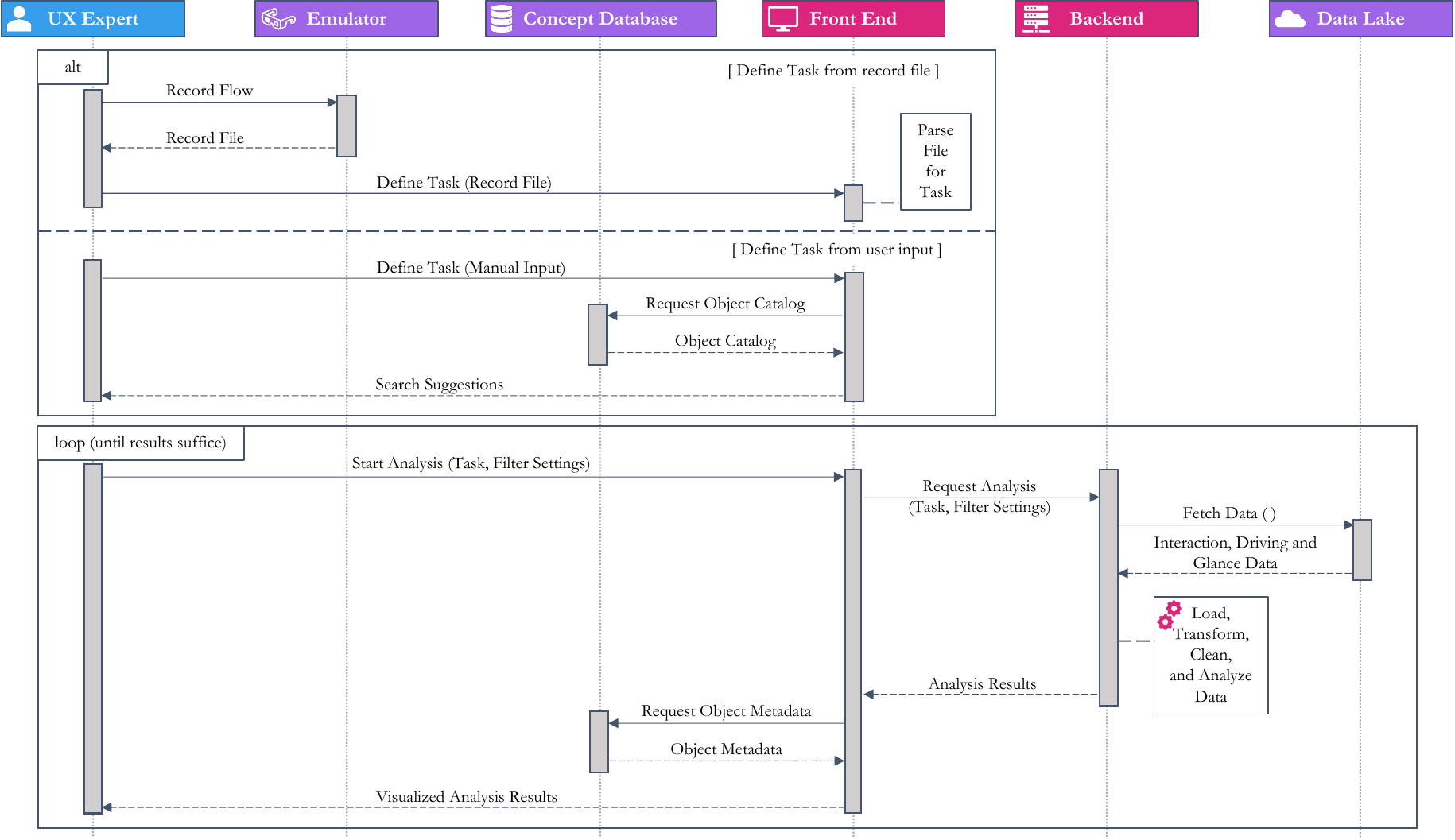}
		\caption{Sequence Diagram showing the interactions between all processes.} 
		\label{fig:SequenceDiagram}
	\end{figure*}

\end{document}